\input psfig

\def\authcomment#1{%
  \gdef\PageFoot{%
    \nointerlineskip%
    \vbox to 22pt{\vfil%
      \hbox to \PageWidth{\noindent {\it #1} \hfil}}%
  }%
}

\def\SA{S97A} 
\def\SB{S97B} 
\def\CHSIX{Sevenster 1997} 

%
%
\def\bck{\hskip-0.35em}
\def\wisk#1{\ifmmode{#1}\else{$#1$}\fi}

\def\oversim#1#2{\lower1.5pt\vbox{\baselineskip0pt \lineskip-0.5pt
     \ialign{$\mathsurround0pt #1\hfil##\hfil$\crcr#2\crcr\sim\crcr}}}
\def\gsim{\wisk{\mathrel{\mathpalette\oversim{>}}}} 
\def\lsim{\wisk{\mathrel{\mathpalette\oversim{<}}}} 

\newcount\levelone    \levelone=0
\newcount\leveltwo    \leveltwo=0
\newcount\levelthree  \levelthree=0
\newcount\levelfour   \levelfour=0
\def\chaphead{}                             
\def\secno{\chaphead\the\levelone}
\def\subno{\chaphead\the\levelone.\the\leveltwo}
\def\subsubno{\chaphead\the\levelone.\the\leveltwo.\the\levelthree}
\def\subsubsubno{\chaphead\the\levelone.\the\leveltwo.\the\levelthree
                           .\the\levelfour}
\def\newsec{\advance\levelone by1 \leveltwo=0 \levelthree=0 \levelfour=0}
\def\newsub{\advance\leveltwo by1 \levelthree=0 \levelfour=0}
\def\newsubsub{\advance\levelthree by1 \levelfour=0}
\def\newsubsubsub{\advance\levelfour by1}
\def\absnarrower{\advance\leftskip by \abstractindent}
\def\titlehang{\hangindent\abstractindent \hangafter 0 \relax}

\newdimen\bottomtol \bottomtol=0.03\vsize
\def\secskip{\par \ifdim\lastskip<\secskipamount \removelastskip \fi
    \vskip 0pt plus \bottomtol \penalty-250
    \vskip 0pt plus -\bottomtol \relax
    \vskip\secskipamount plus3pt minus3pt}
\def\subskip{\par \ifdim\lastskip<\subskipamount \removelastskip \fi
    \vskip 0pt plus 0.5\bottomtol \penalty-150
    \vskip 0pt plus -0.5\bottomtol \relax
    \vskip\subskipamount plus2pt minus2pt}

\outer\def\unnumberedsectionbegin #1\par {\secskip \noindent {\bf #1}
    \nobreak \vskip 6pt \noindent}
\outer\def\sectionbegin #1\par {\secskip \newsec \noindent {\bf \secno.~#1}
    \nobreak \vskip 6pt \noindent}
\outer\def\subsectionbegin #1\par {\subskip \newsub \noindent {\it \subno.~#1}
    \nobreak \vskip 6pt \noindent}
\outer\def\unnumberedsubsectionbegin #1\par {\subskip \noindent {\it #1}
    \nobreak \vskip 6pt \noindent}
\outer\def\subsubsectionbegin #1\par {\subskip \newsubsub \noindent 
    {\rm \subsubno.~#1}
  \nobreak \vskip 6pt \noindent}

\def\aatitle#1\par {{\null \fourteenpoint 
     \vskip 50pt \baselineskip 18pt \titlehang \noindent \bf #1}
     \ifforcopyeditor \vfil \vfil \fi}
\def\aaauthor#1\par {\vskip 16pt \noindent {\bf #1 \par}
     \ifforcopyeditor \vfil \fi}
\def\aainstitution#1\par {\smallskip \noindent {\rm #1 \par}
     \ifforcopyeditor \vfil \vfil \fi}
\def\keywords#1\par {\oneskip {\ifforcopyeditor \narrower \fi 
  \noindent {\bf Key words: \rm #1}}
  \ifforcopyeditor \vskip 0pt plus 10 fil \relax \eject 
     \else \oneskip \hrule height\ruleht \relax \vskip 15pt \fi
}
%
\newcount\notenumber
\notenumber=1
\newcount\eqnumber
\eqnumber=1
\newcount\fignumber
\fignumber=1
\newcount\tabnumber
\tabnumber=1
\newbox\abstr
%
\def\new{{\rm\chaphead\the\eqnumber}\global\advance\eqnumber by 1}
\def\eqref#1{\advance\eqnumber by -#1 \chaphead\the\eqnumber
           \advance\eqnumber by #1 }
\def\?{\eqref{1}}
\def\last{\advance\eqnumber by -1 {\rm\chaphead\the\eqnumber}\advance
     \eqnumber by 1}
\def\eqnam#1{\xdef#1{\chaphead\the\eqnumber}}
\def\appendixbegin#1 #2{\eqnumber=1 \def\chaphead{{#1}}
    \levelone=0\leveltwo=0\levelthree=0\levelfour=0\eqnumber=1\fignumber=1 
    \vskip\subskipamount\noindent{\ninepoint\bf Appendix #1\ \ \ #2}
    \vskip\subskipamount\noindent}
\def\noappendixbegin#1 #2{\eqnumber=1 \def\chaphead{{#1} }
    \levelone=0\leveltwo=0\levelthree=0\levelfour=0\eqnumber=1\fignumber=1 
    \vskip\subskipamount\noindent{}
    \vskip\subskipamount\noindent}
\def\nfig{\chaphead\the\fignumber\global\advance\fignumber by 1}
\def\ntab{\chaphead\the\tabnumber\global\advance\tabnumber by 1}
\def\nfiga#1{\chaphead\the\fignumber{#1}\global\advance\fignumber by 1}
\def\rfig#1{\advance\fignumber by -#1 \chaphead\the\fignumber
            \advance\fignumber by #1}
\def\fignam#1{\xdef#1{\chaphead\the\fignumber}}
\def\tabnam#1{\xdef#1{\chaphead\the\tabnumber}}
%
%

\def\spirnir#1 {, {1996, In: {Minniti, Rix (eds.)
      Spiral galaxies in the NIR}. Heidelberg, p. #1 }}

\def\lodm#1.{, 1986, In: {Israel F.P. (ed.) Light on Dark Matter.}
  Reidel, Dordrecht, p. #1}
\def\lsse#1.{, 1987, In: {Kwok S., Pottasch S.R. (eds.) Late stages of
   stellar evolution.} Reidel, Dordrecht, p. #1}
\def\galaxy#1.{, 1987, In: {Gilmore G., Carswell B.(eds.) Galaxy.}
   Reidel, Dordrecht, p. #1}
\def\torc#1.{, 1988, In: {Fich M.(ed.) Mass of the Galaxy.}
  Toronto University Press, p. #1}
\def\planneb#1.{, 1989, In: {Torres--Peimbert (ed.) Planetary Nebulae.},
  Reidel, Dordrecht, p. #1}
\def\adass#1 {, {1995, In: {Shaw R.A., Payne H.E., Hayes J.J.E. (eds.) PASPC 77,
   Astronomical Data Analysis Software and Systems IV, } p. #1}}
\def\plarin#1.{, 1984, In: {Greenberg R., Brahic A. (eds.) Planetary Rings.},
  Tucson, p. #1}

\def\seng#1 {, {1981, In: { 
      The structure and evolution of normal galaxies}. Cambridge, p. #1 }}

\def\varmic#1 {, {In: {Ferlet, Maillard, Raban (eds.) 
     Variable stars and astrophysical returns of microlensing surveys}. 
     Ed. Fronti\`eres, p. #1 }}

\def\solve#1 {, {1996, In: {Blitz L., Teuben P.(eds.) Proc.
     IAU Symp. 169, Unsolved problems of the Milky Way}.
     Reidel, Dordrecht, p. #1 }}

\def\mosgn#1 {, {1988, In: {Bianchi, Gilmozzi (eds.)
  Mass outflow from stars and Galactic Nuclei.} p. #1 }}
\def\pprg#1 {, {1981, In: {Iben, Renzini (eds.) Physical Processes in
      Red Giants. } p. #1 }}
\def\cbdmw#1 {, {1992, In: {Blitz (ed.) The Center, Bulge and Disk of the
      Milky Way.} Kluwer, Dordrecht, p. #1 }}
\def\planeb#1 {, {1993, In: {Weinberger R., Acker A.(eds.)
      Proc. IAU Symp. 155,
      Planetary Nebulae.} Reidel, Dordrecht, p. #1 }}
\def\mpneb#1 {, {1990, In: {Mennessier M.O., Omont A. (eds.)
      From Miras to Planetary Nebulae. Yvette Cedex: \'Editions
      Fronti\`eres, } p. {#1}\ }}
\def\gents#1 {, {1993, In: {Dejonghe H., Habing H.J. (eds.) Proc.
     IAU Symp. 153, Galactic Bulges}. Reidel, Dordrecht, p. #1 }}
\def\bargal#1 {, {1996, In: {Buta, Crocker, Elmegreen (eds.)
      PASPC 91, Barred Galaxies, } p. #1\ }}
\def\mopste#1 {, {1994, In: Jorgensen U.G. (ed.) Proc. IAU Coll. 146,
     Molecular Opacities in the Stellar Environment. Springer-Verlag, p. #1}}
\def\physr#1 {, {\it Physics Report}{\bf#1},\ }
\def\iauc#1 #2 {, {IAU Circ.\ }{#1, #2}\ }
\def\nature#1 #2 {, {Nat \ }{#1, #2}\ }
\def\science#1 #2 {, {Sci \ }{#1, #2}\ }
\def\aa#1 #2 {, {A\&A}{ #1, #2} }
\def\aal#1 #2 {, {A\&A}{ #1, L#2}\ }
\def\aas#1 #2 {, {A\&AS}{ #1, #2} }
\def\aj#1 #2 {, {AJ\ }{#1, #2}\ }
\def\apj#1 #2 {, {ApJ\ }{#1, #2}\ }
\def\apjl#1 #2 {, {ApJ\ }{#1, L#2 }\ }
\def\apjs#1 #2 {, {ApJS\ }{#1, #2}\ }
\def\araa#1 #2 {, {ARA\&A\ }{#1, #2}\ }

\def\mnras#1 #2 {, {MNRAS\ }{#1, #2}\ }

\def\rpphys#1 #2 {, {Rep. Prog. Phys.\ }{#1, #2}\ }

\def\pasp#1 #2{, {PASP \ }{#1, #2}\ }
\def\qjras#1 {, {QJRAS \ }{#1}\ }
\def\aus#1 #2{, {Aust.~J. Phys.\ }{#1, #2}\ }
\def\actaa#1 {, {Acta Astron.\ }{#1}\ }



\def\refvdV1989{van der Veen 1989}


\def\refvdVH1990{van der Veen \& Habing 1990}














\def\refBS1{ Blitz \& Spergel 1991}




\def\wisk#1{\ifmmode{#1}\else{$#1$}\fi}
\def\fv{\wisk{\, f_{\rm V}}}
\def\mum{\wisk{\mu}m}
\def\kms{\wisk{\,\rm km\,s^{-1}\,}}                    
\def\decdeg#1.#2 {\wisk{#1^{\,\rm o}\bck.\,#2}\ }
\def\losa{line--of--sight}
\def\losn{line of sight}
\def\degr{\wisk{^{\circ}}}                                
\def\decsec#1.#2 {\wisk{#1^{\prime\prime}\hskip-0.42em.\hskip0.10em#2}\ }
\def\kmsr{\wisk{\,\rm km\,s^{-1}\,kpc^{-1}}}
\def\pspeed{\wisk{ \Omega_{\rm p}}}

\def\etal{{et al.$\,$}}
\def\cse{circum--stellar envelope}
\def\cses{circum--stellar envelopes}
\def\df{distribution function}

\def\lvd{longitude--velocity diagram}

\def\rsun{\wisk{\rm \,R_\odot}}
\def\vsun{\wisk{\rm \,V_\odot}}
\def\msun{\wisk{\rm \,M_\odot}}
\def\gbu{galactic Bulge}
\def\gba{galactic Bar}
\def\gc{galactic Centre}

\newcount\fignumber \fignumber=1
\newcount\eqnumber \eqnumber=1

\input mn

\BeginOpening

\title{New constraints on a triaxial model of the Galaxy}

\author{Maartje Sevenster$^{2(1)}$,
 Prasenjit Saha$^{6(2)}$, David Valls--Gabaud$^{3(4)}$, Roger Fux$^5$}

\vskip .5truecm
\affiliation{$^1$Sterrewacht Leiden, 
POBox 9513, 2300 RA Leiden, The Netherlands (msevenst@mso.anu.edu.au)}
\vskip .1truecm
\affiliation{$^2$Mt.Stromlo and Siding Spring Observatories, 
Private Bag, Weston Creek P.O., 2611 ACT Weston, Australia}
\vskip .1truecm
\affiliation{$^3$UMR CNRS 7550, Observatoire de Strasbourg, 
   11 Rue de l'Universit\'e, 67000 Strasbourg, France }
\vskip .1truecm
\affiliation{$^4$Institute of Astronomy, Madingley Road, 
   Cambridge CB3 0HA, UK}
\vskip .1truecm
\affiliation{$^5$Geneva Observatory, Ch. des Maillettes 51, CH-1290 Sauverny,
    Switzerland}
\vskip .1truecm
\affiliation{$^6$Dept. of Physics, Keble Road, Oxford OX1 3RH, UK}

\shortauthor{M.~Sevenster et al.}
\shorttitle{New constraints on a triaxial model of the Galaxy}

\acceptedline{Accepted xx. Received xx}

\abstract{
We determine the most likely values of the free parameters
of an N--body model for the Galaxy developed by Fux
via a discrete--discrete comparison with
the positions on the sky and line--of--sight velocities 
of an unbiased, homogeneous sample of OH/IR stars.
Via Monte--Carlo simulation, we find the plausibilities of
the best--fitting models, as well as the errors on the determined values.
The parameters that
are constrained best by these projected data are
the total mass of the model and
the viewing angle of the central Bar, although the distribution
of the latter has multiple maxima.
The other two free parameters, the size of the Bar
and the (azimuthal) velocity of the Sun, are less well--constrained.
The best model has a viewing angle of $\sim$ 44\degr, semi--major axis
of 2.5 kpc (corotation radius 4.5 kpc, pattern speed 46 \kmsr),
a bar mass of 1.7$\times10^{10}$\msun\
and a tangential velocity of the local standard
of rest of 171 \kms. 
We argue that the lower values that are commonly found 
from stellar data
for the viewing angle ($\sim$25\degr) arise when too
few coordinates are available, when
the longitude range is too narrow or when low
latitudes are excluded from the fit.
The new constraints on the viewing angle of the \gba\
from stellar \losa\ velocities decrease further
the ability of the Bar's distribution to account for the observed
micro--lensing optical depth toward Baade's window :
our model reproduces only half the observed value.
The signal of triaxiality diminishes quickly with increasing
latitude, fading within approximately one scaleheight (\lsim 3\degr). 
This suggests that Baade's window is not a very appropriate region to
sample Bar properties.
}


\maketitle

\section{Introduction}

Evidence has accumulated over the last five years that 
there is a triaxial structure in the inner Galaxy
(see Gerhard 1996 for a review). The study of the \gba\
received a large stimulus when the COBE--DIRBE data and
the derived models became available (Dwek \etal1995).
Earlier, star counts, gas dynamics and three--dimensional
stellar kinematics had been analysed. Although most studies agree
on the presence of a Bar and roughly on its orientation,
the exact viewing angle, size and shape remain a matter of
debate. In this article we use for the first time a global
set of stellar \losa\ velocities in the galactic Plane to determine
the values of these parameters, by comparing the set to a triaxial
N--body model of the Galaxy (Fux 1997).

N--body models are important for
the study of the dynamics of the triaxial Galaxy, because
they are self--consistent and have known 
formation-- and evolution history.
This is opposite to self-consistent Schwarzschild--type
models (Schwarzschild 1979), for which we know only the present and the future.
Schwarzschild--type models can be aimed directly at
fitting observations. True N--body models can much less easily 
be `steered' that way and comparing them with observations is 
difficult. This holds especially for the Galaxy, where stellar 
kinematic data
are always discrete and one is therefore faced with the problem 
of comparing two distributions of discrete data points.
One could smooth and normalize the (projected) N--body model
if it has sufficient particles and sample this probability
distribution at the observed points.
The result is the probability of the data given the model.
One can also smooth the observations to obtain a velocity
profile, as is often done with observations in Baade's window,
and compare that to the corresponding, smoothed, N--body profile.
Statistical tests then yield the probability that the two 
distributions are the same.

We have the opportunity to use a new stellar--kinematic
data set, homogeneous and unbiased, with highly accurate 
on--the--sky positions and line--of--sight velocities
(Sevenster \etal1997a,b; \SA, \SB). 
This data set (hereafter 
AOSP (Australia telescope Ohir Survey of the Plane))
is particularly suited for constraining 
dynamical models for the Galaxy, not only because of its high
accuracy but also because of the intrinsic properties of the stars. 
The AOSP sample consists of so--called OH/IR stars,
observable throughout the Galaxy. These are 
evolved, intermediate--mass stars and their distribution, 
spatial as well as kinematic, traces closely the 
global galactic potential (Habing 1993, Frogel 1988) and
is relatively relaxed. OH/IR stars have \cses\
due to mass loss and the outflow velocity of those \cses\ can
be obtained directly from the spectra. The outflow velocity is 
related to the stellar mass and thus to the 
age of the star, in a statistical sense (van der Veen 1989; 
see \CHSIX). This allows, for example, determination 
of the changes in the dynamical distribution with time.

The average surface density of the sample is of the order of one star
per square degree. This means that two neighbouring stars
cannot be assumed to sample the same velocity profile, 
which is implicitly required to smooth the data.
Also, it is not necessary to smooth
the model completely, because we want to determine the
probability of the model given the data, rather than 
the other way round, or the probability that model and data
have the same distribution. We use a method
to scale an N--body model (Fux 1997) to match the data,
via an implementation of a direct discrete--discrete comparison (Saha 1998).
The model was chosen from a range of N--body models
because it reproduces best the combination of
the COBE--DIRBE surface--density 
map (in the K band (2.2\mum)) and 
other observations (eg. the local dispersions and density, Fux 1997).
It is therefore most representative for the 
AOSP sample, because this comes from exactly the same intermediate--mass,
evolved stellar population that dominates the near infra--red 
surface density observed by COBE. 

In Section 2 we describe briefly the general method 
for the Galaxy model--data 
comparison, in Section 3 we describe the detailed implementation 
for the given data and model we use. The results 
are presented and discussed in Sections 4,5. In Section 6 we
calculate the implications for gravitational micro lensing toward the
\gbu\  and we finish with conclusions in Section 7.

\def\PageFoot{}

\section{The method}

\subsection{Determining the best fit}

To compare the six--dimensional N--body model (cartesian
coordinates $x,y,z,u,v,w$)
and the three-dimensional data (galactic longitude $\ell$ and 
latitude $b$, and \losa\ velocity $V$), the
model is projected according to :

\eqnam\TRANS

 $$ x^{\prime} = x\cos\phi + y\sin\phi\ \  \ \ 
 y^{\prime} = y\cos\phi-x\sin\phi + R_{\odot}   $$

 $$ u^{\prime} = u\cos\phi + v\sin\phi \ \   \ \ 
 v^{\prime} = v\cos\phi-u\sin\phi   $$

 $$ \ell = \arctan \left( {{-x^{\prime} }
    \over{ y^{\prime} }}  \right) \ \ 
  b =  \arctan \left( {z} \over {\sqrt{ [x^{\prime} ]^2 
    + [y^{\prime}]^2 }}  \right) \  $$

  $$ V = f_{V} \{
     {{{ x^{\prime}\, u^{\prime}   +
      y^{\prime}\,v^{\prime} +
      w\,z}}\over
      {{\sqrt{ [x^{\prime}]^2  + 
       [y^{\prime}]^2 + z^2 }}}} \} - 
    V_{\odot} \sin\ell \, \cos b . 
\eqno(\new)  $$

\noindent
The four scaling parameters $\phi, R_{\odot}, f_{\rm V}, V_{\odot}$
are the free parameters of the model. The viewing angle $\phi$ is
the orientation of the Bar with respect to the \losn\ to the
\gc\ (if $\phi=0^{\circ}$ the Bar points toward the Sun). 
$R_{\odot}$ enters equation (\TRANS) as if it were the distance of the
Sun to the \gc, but it determines the size of the Bar ($\ell,b$
become smaller with larger $R_{\odot}$). 
If in the initial model the semi--major axis is $a$, then in the
scaled model it is ($a\times 8\,{\rm kpc}/R_{\odot}$).
This is the consequence of having only ($\ell,b,V$) to fit;
if we had more coordinates there would be an extra free 
parameter $f_{\rm R}$ (the true size scale) and 
\rsun\ would be the true distance to the \gc.
In equation (\TRANS) this parameter $f_{\rm R}$ is hidden, as it
enters the numerator as well as the denominator for 
all three coordinates and thus is of no consequence.
The velocity factor $f_{\rm V}$ determines the total mass,
$M_{\rm p} \propto f_{\rm V}^2\,(8\,{\rm kpc}/R_{\odot})$ (maintaining
virial equilibrium). $V_{\odot}$ is the azimuthal
velocity of the Local Standard of Rest (LSR).
The four free parameters are all weakly correlated, as is clear
from the fact that we have only three quantities to fit. In equation (\TRANS)
we see immediately that the spatial distribution of the data places
no constraints on $f_{\rm V}$ and \vsun. The \losa\ velocities do
constrain \rsun\ which enters the coordinate--transformation
term as well as the solar--motion--correction term in the expression
for $V$. The viewing angle $\phi$ plays a prominent role in all
terms of equation (\TRANS).

We determine the values of the free parameters that optimize 
the model--data fit with a so--called genetic programming
method (Charbonneau 1995). 
We divide the three--dimensional
data--space in $B\equiv N_{\ell}N_{b}N_{V}$ boxes and determine the 
number of model-- and data particles in each box $i$, $m_i$ and
$o_i$, respectively. The joint probability $W$
that the data (in total $O$ particles) 
and the model ($M$ particles), projected on the 
data--space, arise from the
same underlying distribution function, is given by the
following formula (Saha 1998) :

\eqnam\LNW$$
  W =  C \prod_{i=1}^{B} {(m_i + o_i)! \over \left( m_i! o_i! \right)} \ \ \ \ \ , \ \ \
     C = {M! O! (B-1)!\over(M+O+B-1)! } \ \ , $$

  $$ \ \ \ \ \ \ \ \ \    o_i,m_i=o_i,m_i(\phi,R_{\odot},V_{\odot},f_{\rm V}).
\eqno(\new) $$

\noindent
Note that this equation is symmetric in the model-- and data--terms.
$W$ is robust against
outliers in the data (or in the model if $M<O$ which is unlikely
ever to be the case) and also against unphysical solutions, such 
as putting all $M$ model particles in the box with highest $o_i$ (Saha 1998).

If $m_i=0$ or $o_i=0$, no contribution to the likelihood $W$ is made;
the term within the product in equation (\LNW) equals one. 
Preferably, $B < M$ and $B < O$ 
so that we have as few boxes as possible without information content.
On the other hand, we want to prevent any smoothing
of the data, so that $o_i$\lsim 1 for all $i$ and $B > O$.
In other words, $B \sim O$, which results in $B < M$ as $M >>O$.
A fundamental constraint on the box size comes
from the demand that, within a box, the distribution 
function, that $M$ and $O$ derive from, is constant.
This constraint is much harder to quantify in practice, because
we do not know the distribution function.
In general it is also in conflict with the first constraint. 
If we make the boxes so small that the \df\ is truly
constant within each of them, they will not
all contain at least one star.
A proper balance has to be found between the two constraints.
From tests we find that results are robust over a large range
of $B$ -- $2O < B < 8O$. For both larger and smaller $B/O$ the
plausibilities start going to zero, although the best--fit values
remain constant for smaller $B/O$. Judging by the values, biases
and plausibilities of our tests, $B/O \sim 4$ is optimal.

\beginfigure1
\fignam\PROC
{\psfig{figure=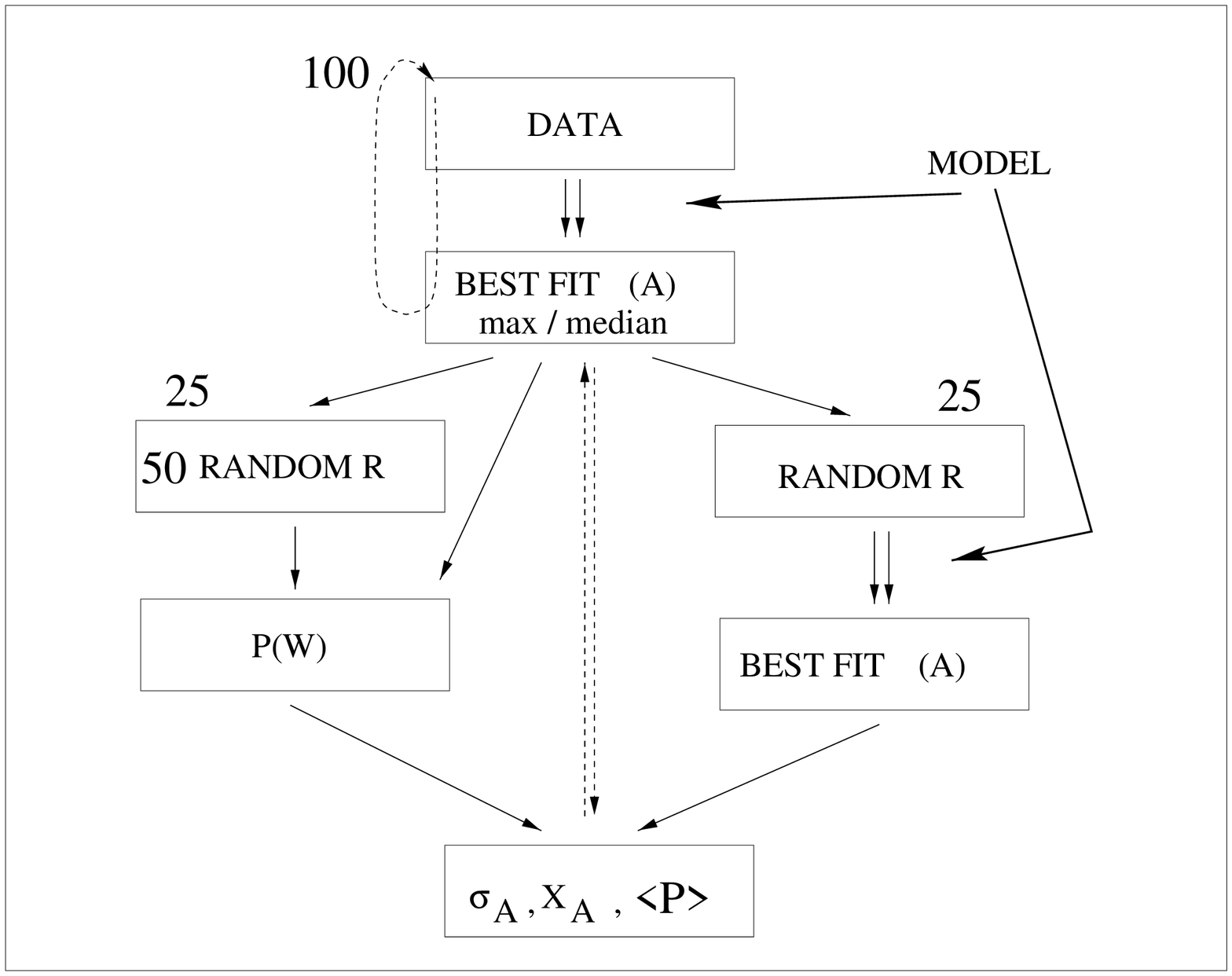,height=6.5cm}}
\caption{{\bf Figure \nfig }
Schematic view of the procedure
to determine the $W$--distribution (left branch) and the 
intrinsic error on the free parameters $A$ (right branch).
The double arrows indicate the steps that involve optimizing (of $W$).
First the maximum of 
$W(\phi,R_{\odot},V_{\odot},f_{\rm V})$ (equation (\LNW))
is determined with a genetic algorithm for 100 different subsets
of the model. From the results, best--fit values $A$ are determined
(see Section 2.1). From the model, scaled
with parameter values $A$, 50 random samples
of $O$ particles are drawn and the value of $W$ for each of 
those ``data sets'' given the scaled model is calculated. 
The whole left branch is executed 25 times, yielding the
plausibility $\langle P\rangle$
(see Section 2.2).
On the other branch, 25 random samples of $O$ particles
are drawn from the scaled model. For each of those, again 
$W(\phi,R_{\odot},V_{\odot},f_{\rm V})$
is optimized, ie. using the initial model. The resulting 
distributions in $\phi,R_{\odot},V_{\odot},f_{\rm V}$ yield
the errors $\sigma$ in and the biases $X$ on the fitted values.
}
\endfigure

In principle all model particles could be used for optimizing
equation (\LNW), but in practice we use random subsets with
$M=16384$, saving calculation time without losing precision.
We carry out the optimization for 100 such subsets to obtain
a number distribution of best--fit values for each parameter (see
Fig.$\,$\PROC).
We then determine the medians of these distributions,
or all local maxima if there are more than one
(only $\phi$ as will be clear later),
thus finding one or more best fits 
($\phi, R_{\odot}, V_{\odot}, f_{\rm V}$).

\subsection{Determining the plausibility}

Having found a {\it best} fit to the data, we want to know
whether it is also a {\it good} fit, within the limits
of the model. Via Monte--Carlo simulation (eg. Press \etal1986) we
determine the ``intrinsic'' $W$ distribution for the best--fit
model. Sets of $O$ model particles are randomly selected
from the (entire) best--fit model and the corresponding value of
$W$ is calculated. The resulting distribution thus gives
the $W$ values for the case we know that the model and
the ``data'' are the same. We then determine the percentage $P(W)$
of these values that is lower than the $W$ of the real
data. A high value of $P(W)$ means that the result is significant;
the fit is as good as can be expected for that particular model. 
We carry out this loop
(the left branch in Fig.$\,$\PROC) 25 times, with 50 different
random subsets each time, to obtain $\langle P\rangle$.
We will call $\langle P\rangle$ the plausibility of the fit. 
Roughly, models with $\langle P\rangle < $ 10\% are not acceptable,
models with $\langle P\rangle >$ 50\% are optimal.

\subsection{Determining the errors}

To quantify the errors that are connected with the determination
of a best--fit model, we again draw
random samples of $O$ particles from the (entire) 
best--fit model and find the
best fit for these fake data sets. The mean value for each of
the free parameters from these fits, and the dispersion in
the values, show the intrinsic accuracy of the fitting procedure. 
For each quantity $A$, we thus find the 1$\sigma$ error 
$\sigma_{\rm A}$ and difference between the best--fit value and the 
Monte--Carlo mean, in terms of the 1$\sigma$ error,
the bias $X_{\rm A} \equiv (A_{\rm mean}-A_{\rm fit})/ \sigma_{\rm A} $. 
In practice, we use 25 independent samples, created in such a way that 
$O \cap M = \emptyset$ .

\section{Implementation}

\subsection{Model and data particles}

The N--body model we use is the model ``m08t3200'' 
(at 3.2 Gyr in the simulation) developed by Fux (1997). 
It contains a bar that formed spontaneously
(without imposed triaxial potential) from an axisymmetric
distribution of stars, in a disk and a spheroid, plus a dark halo.
The stellar part consists of 100,000 particles of
$6.57\times 10^5$\msun.
Corotation is at 5.4 kpc (determined from the moments of inertia)
and the bar's semi--major axis is
3 kpc (out to the start of spiral arms, around the inner ultra--harmonic
resonance) in the initial model. 
The circular velocity, at the radial range in which it is constant,
is 218 \kms, so the pattern speed
of the bar is of the order of 40 \kmsr . The in--plane axis ratio of the model
Bar is approximately 0.5. Symmetry with respect
to the plane was imposed during the simulation.
The ratio of the corotation radius to the semi--major axis is 1.8,
which together with an exponential density profile, makes this
a late--type Bar (Elmegreen 1996, Noguchi 1996).

The AOSP data set consists of 507 OH/IR stars (Section 1) with 
measured on--the--sky positions, accurate to \decsec0.5, and 
\losa\ velocities,
accurate to 1 \kms. The observational errors
are effectively zero in this analysis and will be neglected
in the rest of this article.
The properties of the stars in the AOSP sample (Section 1) allow us to gain more
information from this modelling than just the best fit to 
the full data set. The outflow velocity of the \cse, of the order
of 15 \kms, is roughly proportional to luminosity (van der Veen 1989), 
mass and age
(since these are all the same parameter, given a certain 
stellar--evolutional phase). The relation 
should be applied in a statistical sense. 
Stars with higher outflow velocities can hence be detected out to 
larger distances (\gsim 12 kpc), on average,
than those with lower outflow velocities (\lsim 10 kpc, \CHSIX), 
provided they are observed with the same flux--density cut--off 
(\SA,\SB).
The spectra of some of the stars ($<$20\%) 
show only one of the usual two peaks,
thus not allowing for a determination of the outflow velocity and
an accurate \losa\ velocity (\SA, \SB). For these stars, as a group,
the velocity accuracy is of the order of the average 
outflow velocity, 15 \kms.
We applied the method to the total AOSP sample as well as to 
subsamples, to see how their different properties influence
the fit (Section 3.3).
The \losa\ velocities of the stars are given with respect to
the local standard of rest (LSR, see \SA). Throughout this article, 
we will use \vsun\ and ``solar motion''
to indicate the azimuthal motion of the LSR.

\subsection{The final runs}

\begintable*1
\tabnam\WIND
\caption{{\bf Table \ntab.} Windows for the runs ($M=16384$)}
\tabskip=1em plus 2em minus 0.5em%
\halign to 12cm{
$#$\hfil&\hfil$#$\hfil&\hfil$#$\hfil&\hfil$#$\hfil&\hfil$#$\hfil&\hfil$#$\hfil&\hfil$#$\hfil&\hfil$#$\hfil&\hfil$#$ & # \cr
\noalign{\vskip2pt\hrule\vskip2pt\hrule\vskip2pt}
{\bf Name}&\ell&b&V&N_{\ell}&N_b & N_V & O & B & description \cr
 &^{\circ}&^{\circ}&km\,s^{-1}& & &  & & & \cr
\noalign{\vskip2pt\hrule\vskip2pt}
{\bf bd}& -45.5,10.5&-3.25,3.25&-300,300&20 &6& 15& 500 & 1800 & all \hfil\cr
{\bf b}& -10.5,10.5&-3.25,3.25&-300,300&10 &6& 15& 303 & 900 & bulge \hfil\cr
{\bf bdd}& -45.5,10.5&-3.25,3.25&-300,300&20 &6& 10& 410 & 1200 & double-peaked\hfil\cr
{\bf bdh}& -45.5,10.5&-3.25,3.25&-300,300&20 &6& 10& 250 & 1200 & high $\Delta v$  \hfil\cr
{\bf bdl(l)}& -45.5,10.5&-3.25,3.25&-300,300&20 &6& 10& 250 & 1200& low $\Delta v$  \hfil\cr
{\bf fix*}& -45.5,10.5&-3.25,3.25&-300,300&20 &6& 15& 500 & 1800 & 
   $\phi/R_{\odot}/V_{\odot}/f_{\rm V}$ fixed\hfil\cr
{\bf vel}& -45.5,10.5&-3.25,3.25&-300,300&20 &4& 20& 500 & 1600 & large $N_{\rm V}$ \hfil\cr
{\bf mor}& -45.5,10.5&-3.25,3.25&-300,300&40 &4& 10& 500 & 1600 & large $N_{\ell}$ \hfil\cr
{\bf morx}& -45.5,10.5&-3.25,3.25&-300,300&100 &15 & 1& 500 & 1500& $N_{\rm V}=1$ \hfil\cr
{\bf ptl}& -25.,25.&-5.,5.&-250,250&20 &5& 8 & 225 & 800& IRAS based \hfil\cr
\noalign{\vskip2pt\hrule\vskip2pt}
}
\endtable

\noindent
The outflow velocity of 
OH/IR stars correlates, as mentioned above, with average distance.
We used that property
by running the fitting program with various
subsets of the data.
In Table \WIND\ the windows on the sky, the number of boxes in
each coordinate, the number of stars $O$
and the total number of boxes $B$
that we used in the various runs are given.
For all runs, $M=16384$ so that the fits do not depend on
the number of model particles within the window.

As mentioned in Section 2.1, optimally $B/O \sim 4$.
The ``standard'' run {\bf bd} uses the total AOSP sample and the
total region on the sky. Setting
($N_{\ell},N_{b},N_{\rm V}$)=(20,6,15) was found optimal 
in tests for this run. They correspond to bin sizes
of \decdeg2.75, \decdeg1.1 and 40\kms .
For the other runs, we try to stick to those values of $B/O$
and the bin sizes as closely as possible given the
individual run's characteristics.

In run {\bf b}, we use the largest possible longitude
range symmetric about the \gc.
In run {\bf bdd}, we use only sources with double--peaked spectra,
because they have the best--defined \losa\ velocities, with, 
for this application, negligible errors (1 km/s, \SA, \SB).

In run {\bf bdl} we use only sources with
outflow velocities between 1\kms and 15\kms\
and in run {\bf bdh} only those with outflow velocities larger than 13\kms.
The ranges in outflow velocities overlap slightly
in order to retain sufficient stars for the comparison.

The {\bf bdh} contains stars at large distances (Section 3.1)
and the fit for this run 
will be sensitive to the full morphology of the Bar.
For run {\bf bdl}, the data particles and model particles
probably do not trace the same distances; the subsample is
complete out to \lsim 10 kpc
(Section 3.1). We therefore have run {\bf bdll}
with a distance window (0 kpc - 10 kpc) for the model.

To assess the degree to which the fit is determined by the
stellar velocities or by their positions, 
we ran the program with increased velocity-- ({\bf vel})
and spatial resolution ({\bf mor}), respectively.
Those two runs have the same $B$ and $O$ to facilitate comparison.
We also ``switched off'' the kinematics completely ({\bf morx}),
as an extreme test, keeping in mind that $N_{\rm V}=1$ does not
satisfy the demands on the box sizes.
Finally, to judge better the suitability of the AOSP sample for 
constraining the free parameters of the N--body model, 
in {\bf ptl} we used a different sample of
OH/IR stars (te Lintel Hekkert \etal1991).
This sample is incomplete in a number of ways. First, it is
an IRAS--selected sample, which means it is incomplete at
very low latitudes where confusion limited the number of
point sources the IRAS satellite detected. Second, the velocity
coverage changes with longitude. Third, the infra--red selection
(see te Lintel Hekkert \etal1991) introduces a inhomogeneous
distance sampling that is difficult to quantify.
The reason for using this sample nevertheless, or actually 
because of all this, will become clear. 

In {\bf fixf} we fix all parameters, except $\phi$, at
the values found by Fux (1997) for the model m08t3200; 
the value used for \vsun\ is that of the local circular velocity
in his best--fit model. Furthermore, in {\bf fix1} we fix
$\phi$ at 25\degr, in {\bf fix2} at 45\degr\ and in in {\bf fix3}
at 65\degr, all three maxima in the distributions
of best--fit values of $\phi$ in various runs.

\subsection{The search ranges}

In preliminary tests, we found that the optimal search ranges are
($0^{\circ}-90^{\circ}$) for
$\phi$, (6 kpc - 10 kpc) for \rsun, (160\kms - 230 \kms) for \vsun\ and
(0.25 - 2.25) for $M_{\rm p}$. Since scaling the total mass 
means scaling the potential and hence
the velocities squared, in practice we determine
this parameter by scaling a velocity factor $f_{\rm V}$
between (0.5 - 1.5), $M_{\rm p}\propto f^2_{\rm V}$.
The only exception is run {\bf morx}, where we use 
(0.3 - 3.0) for  $f_{\rm V}$.
For the observable quantities, these ranges span amply the likely real
values.

\section{Results }

\beginfigure2
\fignam\BDDW
{\psfig{figure=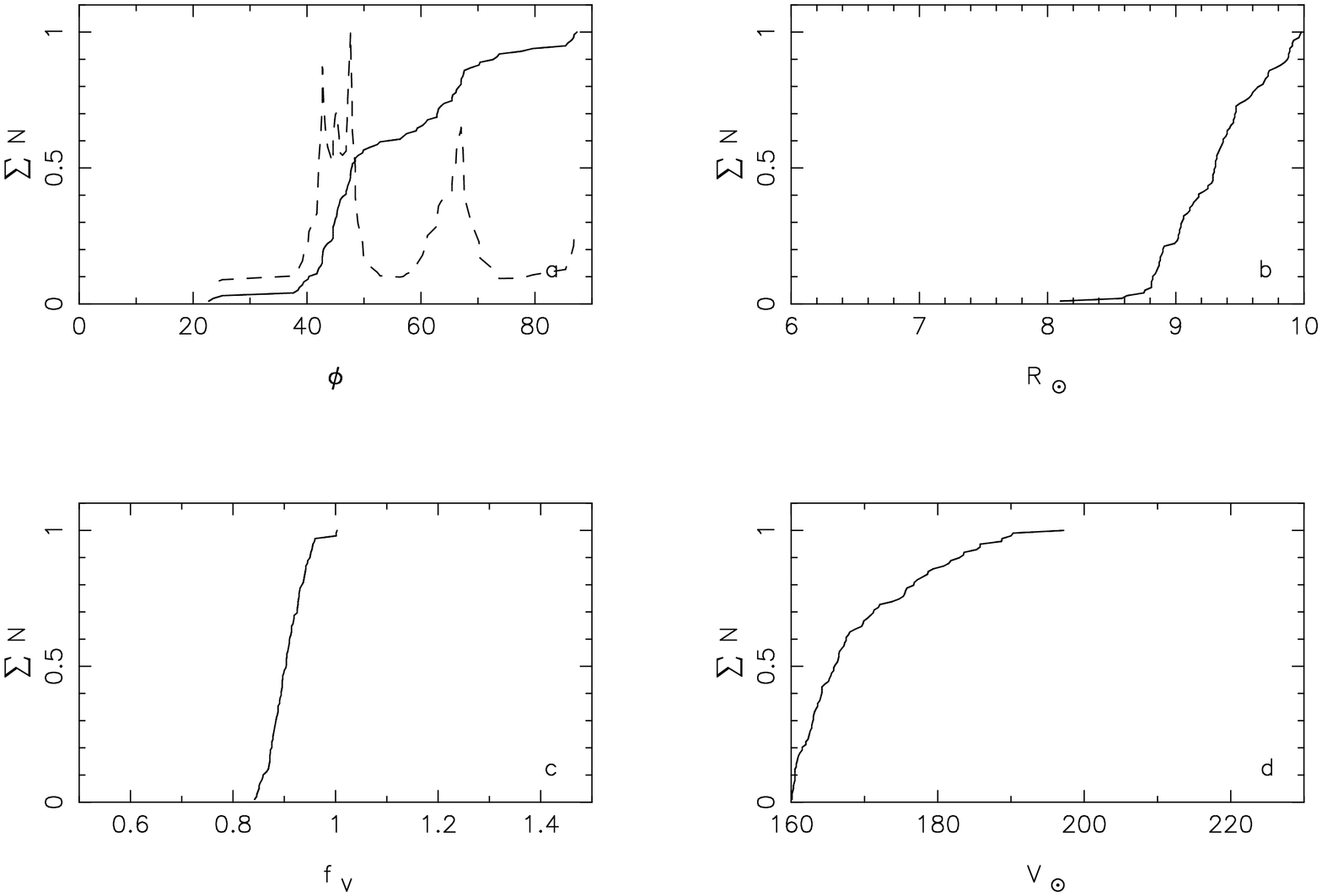,angle=270,width=8truecm}}
\caption{{\bf Figure \nfig}
The cumulative number distributions from the 100 best--fit 
values for the four free parameters from run {\bf bdd}.
For $\phi$ (panel {\bf a}), the dashed curve gives the
(arbitrarily normalized) derivative of the 
cumulative number distribution and hence shows the 
(unbinned) actual number distribution.
}
\endfigure

\beginfigure3
\fignam\BDLW
{\psfig{figure=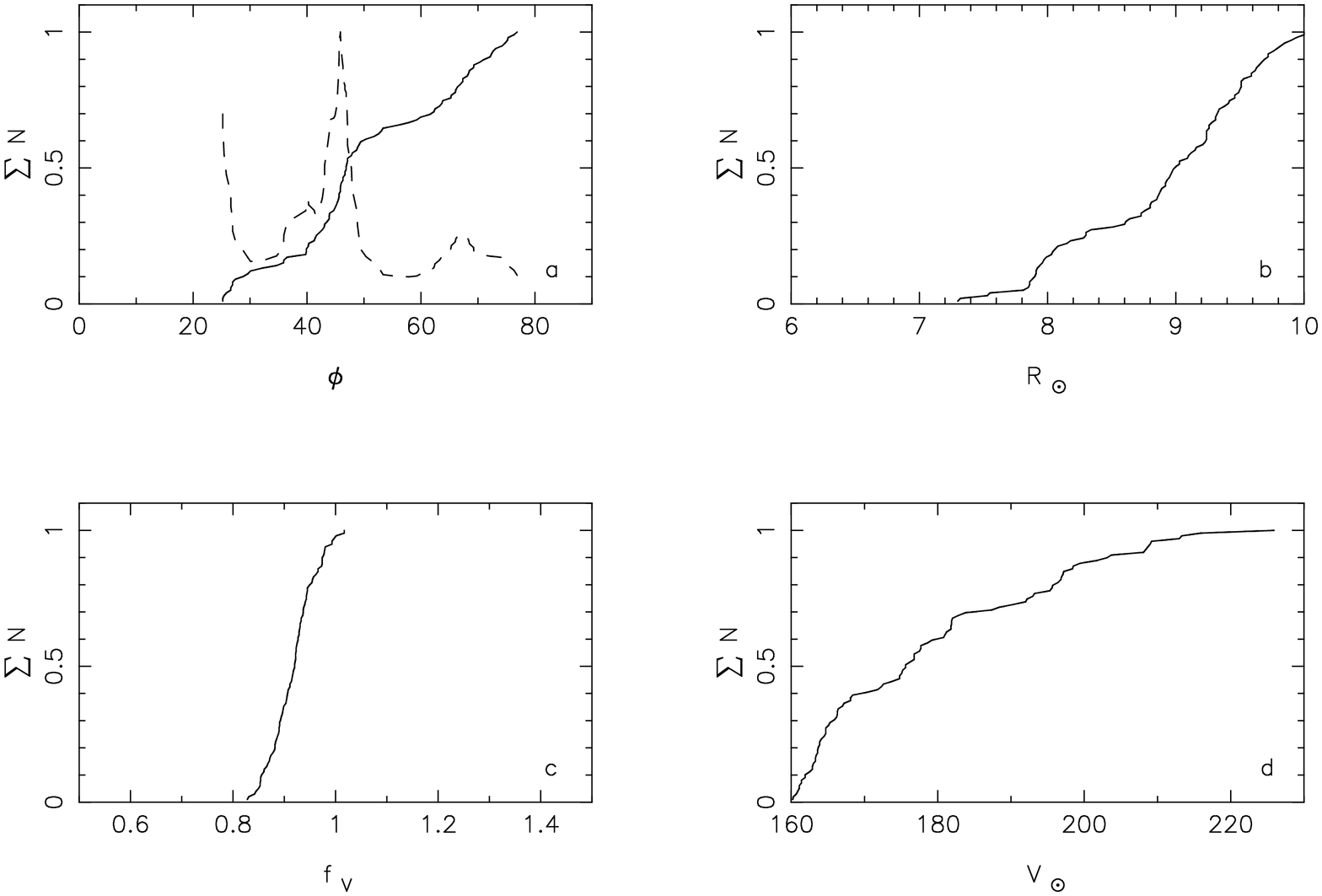,angle=270,width=8truecm}}
\caption{{\bf Figure \nfig}
As Fig.$\,$\BDDW, for run {\bf bdl}.
}
\endfigure

\beginfigure4
\fignam\BDLLW
{\psfig{figure=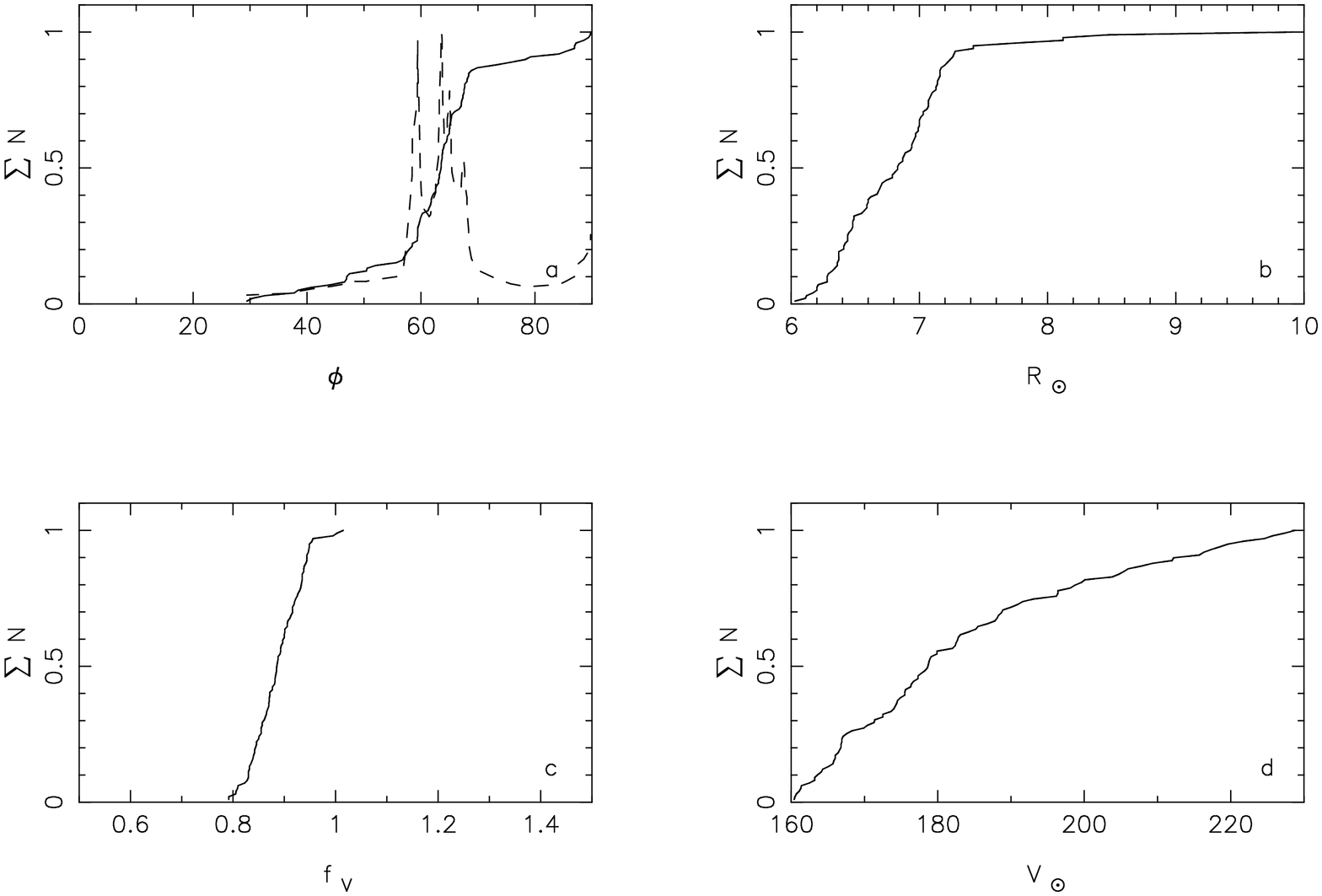,angle=270,width=8truecm}}
\caption{{\bf Figure \nfig}
As Fig.$\,$\BDDW, for run {\bf bdll}.
}
\endfigure

\beginfigure5
\fignam\PTLW
{\psfig{figure=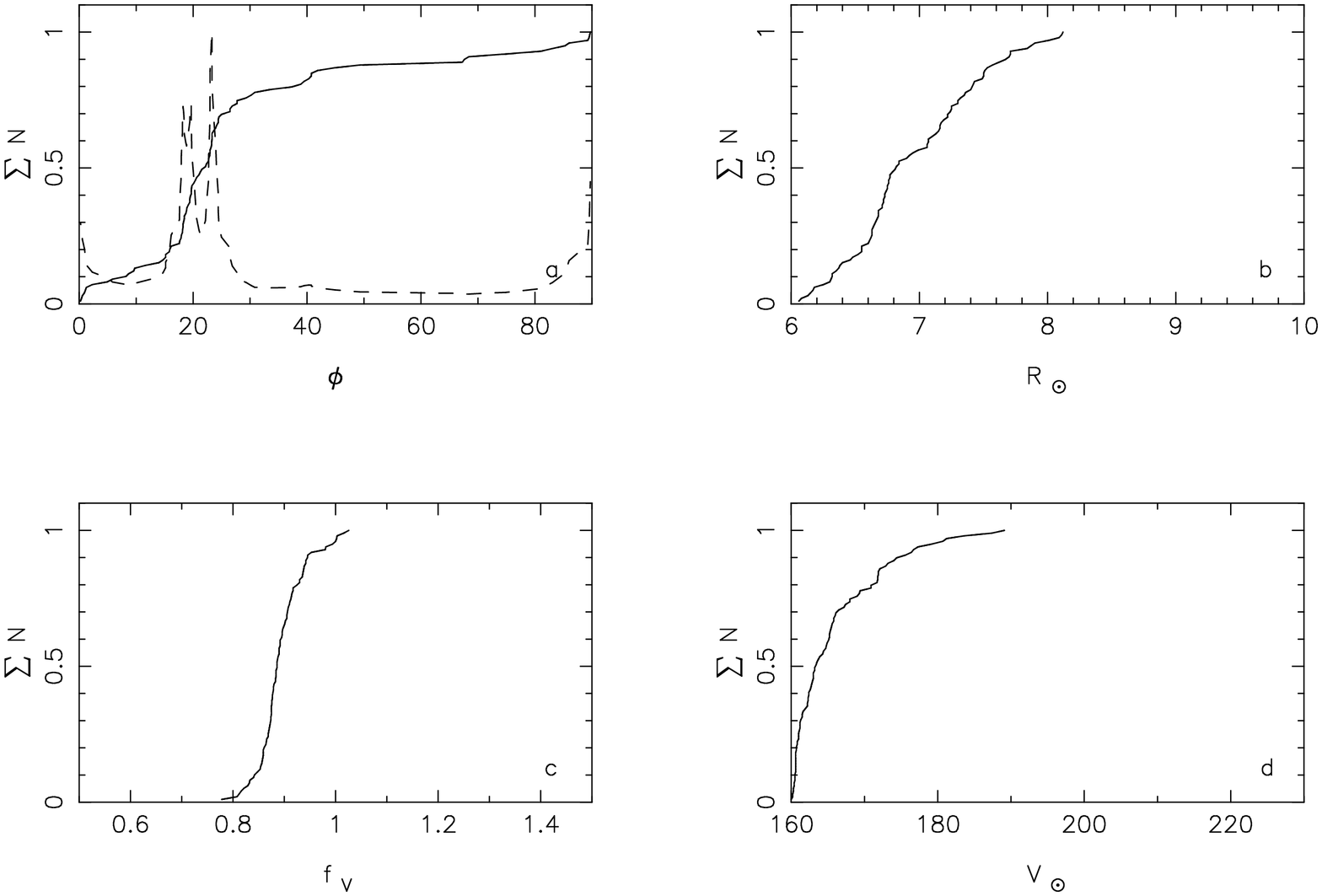,angle=270,width=8truecm}}
\caption{{\bf Figure \nfig}
As Fig.$\,$\BDDW, for run {\bf ptl}.
}
\endfigure

\noindent
Figures \BDDW --\PTLW\ show the results of the 100 $W$--optimizations
(see Fig.$\,$\PROC ) for some of the runs. 
The values of the free parameters are found to be
virtually uncorrelated, so that we can determine the maxima
or medians of
these distributions for each parameter separately to obtain
the best--fit models ($\phi,R_{\odot},V_{\odot},f_{\rm V}$).
Clearly, the distributions of 
$R_{\odot}$ and $V_{\odot}$ do not have well--defined maxima
(inside the search range). We use the median value for those
parameters, as well as for $f_{\rm V}$. The latter's distribution has
the best--defined maximum of all four parameters
(except in {\bf morx}) but owing to its symmetric and
smooth distribution the maximum is the same as the median value in all
cases. We find several clear local maxima only for $\phi$ and 
determine these from the derivatives of the cumulative
number distributions (ie. the unbinned actual number distributions; 
see the dashed curves in Figs$\,$\BDDW --\PTLW ).

\begintable*2
\tabnam\BESTFF
\caption{{\bf Table \ntab.} Best--fit
parameter values (local maxima for $\phi$, medians 
 for $R_{\odot},\,M_{\rm p},\,V_{\odot}$) and plausibilities}
\tabskip=1em plus 2em minus 0.5em%
\halign to 16cm{
$#$\hfil&\hfil$#$\hfil&\hfil$#$\hfil&\hfil$#$&\hfil$#$\hfil
 &\hfil$#$\hfil&\hfil$#$&\hfil$#$\hfil&\hfil$#$\hfil&\hfil$#$
 &\hfil$#$\hfil&\hfil$#$\hfil&\hfil$#$&\hfil$#$\hfil&\hfil$#$\hfil& # \hfil\cr
\noalign{\vskip2pt\hrule\vskip2pt\hrule\vskip2pt}
{\bf Name}&\phi&\sigma_{\phi}&X_{\phi}\ & R_{\odot}&\sigma_{\rm R}
&X_{\rm R}\ & f_{\rm V}&\sigma_{\rm f}&X_{\rm f}
 &\  V_{\odot}
 &\sigma_{\rm V}&X_{\rm V}&\ \langle P\rangle & \sigma_{\rm P} & description \cr
 &^{\circ}&^{\circ}& &\rm kpc&\rm kpc& & & & &\rm km/s&\rm  km/s& & \% & \% & \cr
\noalign{\vskip2pt\hrule\vskip2pt}
{\bf bd (1)}& \bf 44 & 15 & -0.35 &\bf  9.5 & 0.5 & -0.15 
  &\bf  0.95 & 0.04 & -0.20 & \bf 171 & 16 & 0.6 & 15 & 7 & all \hfil\cr
{\bf }& 59 & 20 & -0.02 & 9.5 & 0.4 & -0.03
  & 0.95 & 0.04 & -0.21 & 171 & 18 & 0.4 & 28 & 10 &  \hfil\cr
{\bf }& 71 & 16 & 0.08 & 9.5 & 0.5 & -0.28
  & 0.95 & 0.04 & -0.11 & 171 & 16 & 0.5 & 15 & 8 &  \hfil\cr
\noalign{\vskip2pt\hrule\vskip2pt}
{\bf b}& 55 & 25 & -0.58 & 9.7 & 1.3 & -0.79
  & 0.99 & 0.08 & -1.00 & 212 & 18  & -0.9 & 52 & 12 & bulge \hfil\cr
{\bf }& 60 & 19 & -0.25 & 9.7 & 0.5  & -0.90
  & 0.99 & 0.09 & -0.03 & 212 & 21  & -0.7 & 60 & 13 &  \hfil\cr
{\bf }& 68 & 18 & -0.00 & 9.7 & 0.6 & -0.79
  & 0.99 & 0.06 & -0.83 & 212 & 24  & -1.0 & 56 & 11 &  \hfil\cr
\noalign{\vskip2pt\hrule\vskip2pt}
{\bf bdd}& 43 & 21 & -0.06 & 9.3  & 0.7 & -0.34
  & 0.90 & 0.03 & -0.03 & 166 & 13 & 0.5 & 54 & 10 & double-peaked \hfil\cr
{\bf }& 48 & 19 & -0.54 & 9.3  & 0.6 & -0.08
  & 0.90 & 0.05 & -0.02 & 166 & 15 & 0.7 & 57 & 10 &  \hfil\cr
{\bf }& 67 & 13 & 0.33 & 9.3 & 0.6  & 0.16
  & 0.90 & 0.04 & 0.19 & 166 & 16 & 0.8 & 57 & 10 &  \hfil\cr
\noalign{\vskip2pt\hrule\vskip2pt}
{\bf bdh}& 45 & 20 & -0.43 & 9.6 & 0.4 & -0.41
  & 0.90 & 0.06 & -0.04 & 164 & 7 & 1.1 & 71 & 7 & high $\Delta v$ \hfil\cr
{\bf }& 61 & 24 & -0.36 & 9.6 & 0.5 & -0.59
  & 0.90 & 0.06 & 0.15 & 164 & 15 & 0.8 & 66 & 9 &  \hfil\cr
{\bf }& 83 & 14 & -1.11 & 9.6 & 0.5 & -0.49
  & 0.90 & 0.06 & 0.27 & 164 & 15  & 0.7 & 57 & 9 &  \hfil\cr
\noalign{\vskip2pt\hrule\vskip2pt}
{\bf bdl}& 25 & 18  & -0.07 & 9.0 & 0.7 & 0.09
 & 0.92 & 0.05 & 0.15 & 176 & 16 & 0.2 & 17 & 8 & low $\Delta v$ \hfil\cr
{\bf }& 46 & 22  & -0.16 & 9.0 & 0.7 & 0.28
  & 0.92 & 0.08 & 0.09 & 176 & 22 & 0.4 & 52 & 8 & \hfil\cr
{\bf }& 67 & 14  & -0.22 & 9.0 & 0.8  & -0.11
  & 0.92 & 0.07 & -0.33 & 176 & 17 & 0.1 & 43 & 10 &  \hfil\cr
\noalign{\vskip2pt\hrule\vskip2pt}
{\bf bdll}& 59 & 19 & 0.25 & 6.8 & 0.7 & 0.66 
  & 0.89 & 0.06 & -0.39 & 179 & 21 & 0.2 & 30 & 9 & low $\Delta v, d_{lim}=10$ \hfil\cr
{\bf }& 64 & 15 & -0.13 & 6.8 & 0.8 & 0.48
  & 0.89 & 0.06 & -0.10 & 179 & 17 & 0.8 & 34  & 9 &  \hfil\cr
\noalign{\vskip2pt\hrule\vskip2pt}
{\bf vel}& 44 & 18 & -0.09 & 9.5 & 0.6 & -0.36
  & 0.94 & 0.06 & 0.10 & 169 & 15 & 0.5 & 10 & 5 & large $N_{\rm V}$  \hfil\cr
{\bf }& 56 & 21 & -0.24 & 9.5 & 0.6 & -0.15
  & 0.94 & 0.05 & -0.53 & 169 & 14 & 0.4 & 7 & 5 & \hfil\cr
{\bf }& 64 & 21 & -0.30 & 9.5 & 0.6 & -0.48
  & 0.94 & 0.03 & -0.53 & 169 & 14 & 0.4 & 13 & 8 & \hfil\cr
\noalign{\vskip2pt\hrule\vskip2pt}
{\bf mor}& 41 & 16 & -0.26 & 8.9 & 0.7 & 0.44 
  & 0.92 & 0.05 & 0.03 & 166 & 14 & 0.8 & 18 & 7 & large $N_{\ell}$ \hfil\cr
{\bf }& 54 & 16 & -0.11 & 8.9 & 0.6 & 0.46 
  & 0.92 & 0.04 & 0.05 & 166 & 18 & 0.7 & 25 & 10 & \hfil\cr
{\bf }& 58 & 20 & 0.01 & 8.9 & 0.6 & 0.36
  & 0.92 & 0.04 & -0.78 & 166 & 9 & 0.7 & 28 & 9 & \hfil\cr
{\bf }& 62 & 14 & 0.46 & 8.9 & 0.6 & 0.25
  & 0.92 & 0.05 & 0.40 & 166 & 18 & 0.9 & 25 & 10 & \hfil\cr
\noalign{\vskip2pt\hrule\vskip2pt}
{\bf morx}& 29 & 17 & 0.03 & 9.0 & 0.8 & -0.51
 &\it 1.65 &\it 0.81 &\it 0.22 &\it 200 &\it 22 &\it -0.5  & 0 & 0 & 
   $N_{\rm V}=1$ (large $f_{\rm V}$ range) \hfil\cr
{\bf }& 52 & 20 & 0.11 & 9.0 & 1.0 & -0.68
  &\it 1.65 &\it 0.78 &\it -0.14 &\it 200 &\it 21 &\it -0.5 & 12 & 10 & \hfil\cr
\noalign{\vskip2pt\hrule\vskip2pt}
{\bf ptl}& 18 & 17 & 0.05 & 6.8 & 1.0 & 0.99
  & 0.89 & 0.06 & 0.24 & 163 & 20 & 1.7 & 0 & 1 & IRAS based \hfil\cr
{\bf }& 23 & 17 & 0.20 & 6.8 & 0.8 & 0.78 
  & 0.89 & 0.05 & 0.07 & 163 & 21 & 1.3 & 0 & 1 &  \hfil\cr
\noalign{\vskip2pt\hrule\vskip2pt}
{\bf fixf}& 39 & 17 & -0.30 & 9.0 & - & -
  & 0.98 & - & - & 214 & - & - & 27 & 11 & 
     fixed $R_{\odot}, V_{\odot}, f_{\rm V}$  \hfil\cr
{\bf }& 44 & 20 & -0.22 & 9.0 & - & -
  & 0.98 & - & - & 214 & - & - & 35 & 12 & \hfil\cr
{\bf }& 52 & 18 & -0.14 & 9.0 &  - & - 
  & 0.98 & - & - & 214 & - & -  & 38 & 11 & \hfil\cr
{\bf }& 67 & 19 & 0.08 & 9.0 & -  & - 
  & 0.98 & - & - & 214 & - & -  & 34 & 9 & \hfil\cr
\noalign{\vskip2pt\hrule\vskip2pt}
{\bf fix1}& 25 &  - & -  & 9.4 & 0.5 & -0.04 
  & 0.95 & 0.05 & -0.25 & 172 & 11 & 0.1 & 3 & 3 & 
     fixed $\phi$ (large $f_{\rm V}$ range) \hfil\cr
{\bf fix2}& 45 &  - &  - & 9.3 & 0.7 & 0.18 
  & 0.98 & 0.05 & -0.13 & 168 & 12 & 0.8 & 45 & 11 & 
     fixed $\phi$ (large $f_{\rm V}$ range)  \hfil\cr
{\bf fix3}& 65 &  - & -  & 9.8 & 0.4 & -0.45
  & 0.95 & 0.05 & -0.51 & 173 & 12 & 0.1  & 12 & 6 & 
     fixed $\phi$ (large $f_{\rm V}$ range)  \hfil\cr
\noalign{\vskip2pt\hrule\vskip2pt}
}
\endtable

\noindent
In Table \BESTFF, we give the sets of best--fit values 
for each of the runs described, along with 
the plausibility $\langle P\rangle$ of this best fit
and the spread in the plausibility $\sigma_{\rm P}$.
For each of the parameters, we also give
the 1$\sigma$ error, $\sigma_{\rm A}$, and
the difference between the best--fit value and the
bias $X_{\rm A}$, determined as described in Section 2.
It should be noted that for 
best--fit values at an upper or lower boundary of a search 
range, $X_{\rm A}$ is necessarily large, negative or positive, respectively.
In run {\bf morx} without the kinematics, the parameters \vsun\
and \fv\ (in italics) are not constrained at all.

\beginfigure6
\fignam\PHIS
{\psfig{figure=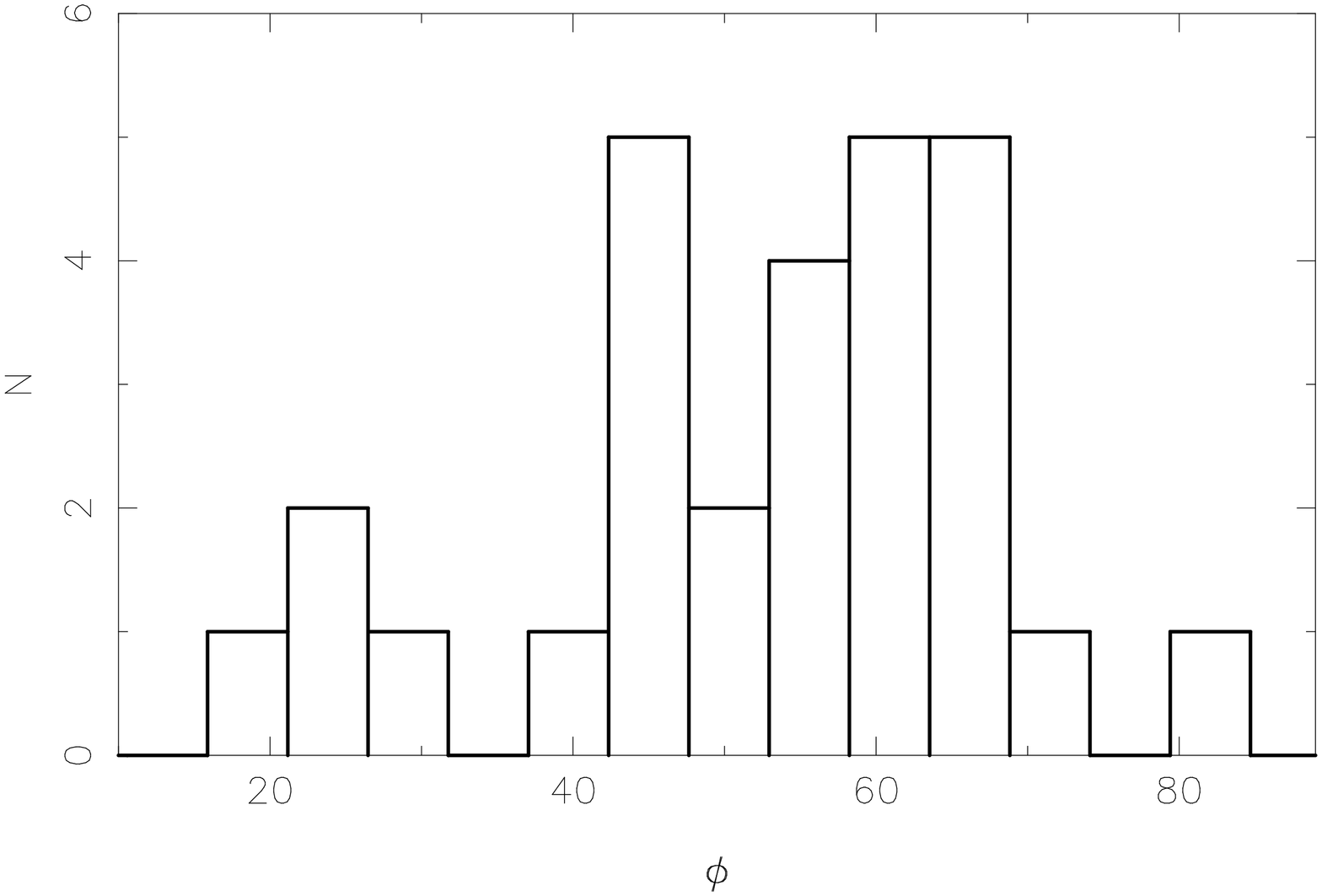,angle=270,width=8truecm}}
{\psfig{figure=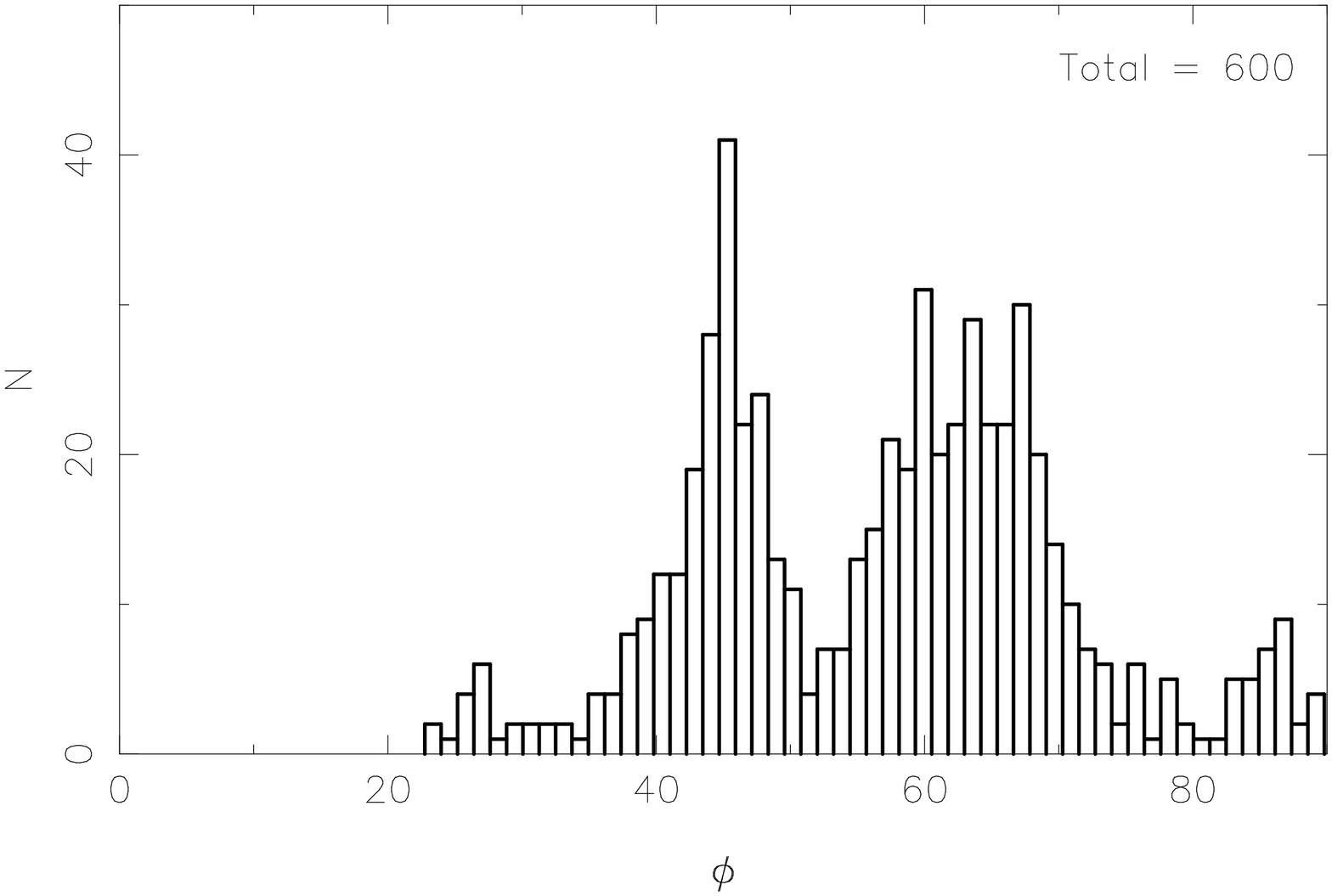,angle=270,width=8truecm}}
\caption{{\bf Figure \nfig} In the top panel, the histogram of
all best--fit values of $\phi$ (from Table \BESTFF ) is shown; in the bottom
panel of all 100 $W$--optimizing values from the six {\bf bd*} runs.
Note the narrow peak at 45\degr, the wide peak around 65\degr\ and
the small peak around 25\degr, as well as the absence of values 
below 20\degr. The peak at 85\degr\ probably indicates 
near--symmetry (90\degr ) of some data sets.
} 
\endfigure

In general, it is clear that $R_{\odot}$ and $V_{\odot}$ are
not optimally constrained by our method and/or data. Their
best--fit distributions do not show clearly isolated maxima and
from Table \BESTFF\ we see that these parameters have 
average biases of $\sim$0.5, as opposed to $\sim$0.2 for $\phi$ and
$f_{\rm V}$. In terms of the ratio of $\sigma_{\rm A}$
to the corresponding search range, $f_{\rm V}$ is very well
confined with $\sigma_{\rm f}$ on average 5\% of the search range.
For the three other parameters this ratio is 20--25\%.
There is good agreement to within $\sim1\sigma$ between the various models
(except {\bf b, bdll, morx} and {\bf ptl}, this will be discussed
later) on the values of the parameters $R_{\odot}$ 
(8.9--9.6), $V_{\odot}$ (164--179) and $f_{\rm V}$ (0.90--0.95).
For $\phi$ the situation is considerably different.
Interestingly, despite the fact that the best--fit distribution
of $\phi$ mostly has several maxima, the subsequent Monte--Carlo
analysis of each of the solutions shows that some are remarkably
well confined. In Fig.$\,$\PHIS\ we show the histograms of the
values for $\phi$ in Table \BESTFF\ as well as of all the
values for $\phi$ occurring in the 100 optimizations for
the six {\bf bd*} runs. The three peaks around 
25\degr, 45\degr\ and 65\degr obviously instigated the runs {\bf fix1--3}.
We disregard the fourth peak at 85\degr\ as this really indicates an
axisymmetric solution. For $\phi$=90\degr\ (side--on) the surface--density
profile is completely symmetric.

In principle run {\bf bd}, 
using the most datapoints and the largest longitude window,
should give the best results. Of its three solutions, the
44\degr\ fit coincides most closely with fits from
several other runs (Fig.$\,$\PHIS). 
Most noteably, the runs with increased
velocity accuracy, either from the data ({\bf bdd}) or from the
method ({\bf vel}), have a solution for similar $\phi$ with very
small bias in $\phi$. Also {\bf bdh}, that samples completely
all distances throughout the Bar, has a fit with $\phi$=45\degr.
These four $\phi\sim$44\degr\ runs 
give similar values for \rsun\ and \vsun, and reasonable agreement
for \fv . Moreover, the {\bf bdh} $\phi$=45\degr\ fit has very high
plausibility, as has {\bf fix2} with respect to {\bf fix1,3}.
The $\phi$=44\degr\ result for {\bf bd} (to be called {\bf bd1}) therefore
gives the best scaling parameters for Fux's N--body model.

\subsection{The degeneracy in $\phi$}

Obviously, there are other values of $\phi$ that give
equally reasonable fits (judging by the errors, biases and
plausibilities). In the lower panel of Fig.$\,$\PHIS\ we
see that $\phi$ even follows an intrinsically quadrumodal
distribution ! It was shown by Zhao (1997b) that,
from surface density only, the viewing angle cannot be 
determined uniquely. Different density models for
the Bar can give exactly the same projected densities
with different viewing angles. It is expected, however, that
no such degeneracy would exist when optimizing a given spatial 
density model, or when including global kinematics 
constraining self--consistent model dynamics. 
The fact that we still find a degeneracy is probably 
at least partly dictated by the limited extent of our longitude
window at positive longitudes. `Asymmetry profiles'
(positive--to--negative--longitude ratios of 
surface density or mean velocity etc.) are crucial in 
constraining $\phi$ and these profiles are cut short.
In the next paragraph we discuss what 
influence this may have. 

\vskip .5truecm
In runs {\bf ptl} and {\bf bdll} \rsun\ is $\sim$2$\sigma$
below the value of the other runs. The samples used in
both runs have larger (apparent) scaleheights than 
the whole population of evolved stars; in {\bf ptl} because 
the sample is incomplete in the plane 
and in {\bf bdll} 
because the distribution of older stars simply has 
larger (intrinsic) scaleheight. 
Run {\bf bdl}, however, the equivalent of {\bf bdll} but without
integration limit (see Section 3.1), results in an 
average value for \rsun. Possibly, 
this can be explained by the facts that we are
not using the integration limit and
that the observed longitude range extends to $+$10\degr, ie.
the near tip of the Bar is outside the window, as we mentioned before. 
This has a two--fold effect. 
First, the surface--density asymmetry profile
of the cut--off bar may look like the profile of an entire bar
with larger viewing angle. We suggest that this is the origin
of the $\phi$ values around 25\degr, as these are seen in
{\bf bdl, morx} and {\bf ptl}.
All samples that contain the {\bf bdl} stars are clearly 
incomplete at larger distances to some extent, most of all
{\bf ptl} (the integration limit can not be
estimated for this sample). Apparently, the kinematics 
suppress the tendency for $\phi$ to be $\sim$25\degr\ (see also
Fig.$\,$\PHIS) if the
incompleteness is not too strong, 
so that we only see it in run {\bf morx}, that does not include
kinematics, and in {\bf bdl} and {\bf ptl} with severe
incompleteness (not taken into account.
Second, one can imagine that 
with the far end of the Bar `cut off' in the data by the
flux density limit and the near end by the longitude limit,
the result may look like a smaller bar, in other words 
: larger \rsun. This may cause the different \rsun\ in 
{\bf bdl} and {\bf bdll}. One would expect the larger
\rsun\ in {\bf ptl} if it didn't have the longitude extend
up to $+$25\degr.

Just as {\bf bdll} and {\bf ptl} have larger than 
average scaleheight, {\bf bdh} has smaller scaleheight.
Indeed, in {\bf bdh} the value for \rsun\ is relatively 
large. The deviation is not as large as for {\bf bdl}, as the
{\bf bdh} sample spans the whole Bar and puts stringent
constraints on the surface--density asymmetry profile.

\vskip .5truecm
\noindent
The values for \vsun\ are rather low compared to the
200$\pm$20 \kms\ currently accepted 
(see Dehnen \& Binney 1998; Feast \& Whitelock 1997; 
Rohlfs \etal1986). However, these determinations all
assume \vsun\ is the local circular speed which is
unlikely, amongst other reasons because the disk may be slightly elliptical
(Kuijken \& Tremaine 1994). These authors advocate an average
local circular velocity of 200 \kms\ and a tangential
velocity for the LSR of 180 \kms, making the low fitted
values for \vsun\ more acceptable. The difference with
the model's circular velocity of 207 \kms\ is still large.
In run {\bf b} \vsun\ is particularly ill--determined,
because the $\sin \ell$ term in the correction for the \losa\ velocities
(equation (\TRANS)) covers a small range only.

Very high velocities (up to 450\kms)
are present in the initial N--body model that were never
found for OH/IR stars. Altogether,
less than ten stars are known at absolute 
velocities higher than 300 \kms\ (Baud \etal1975; 
van Langevelde \etal1992; \SA). 
These high model velocities are not used in the
comparison, however, 
because they are outside the windows defined in Table \WIND.
Those windows are the true limits of the observations (\SA, \SB) so
increasing the velocity window would be meaningless. 
The data--model comparison, in particular the determination of
\fv, is therefore not based on the extreme tails,
but on the wings of the distribution of the bulk of the velocities.
The resulting \fv $<$1 shows that
the total mass $M_{\rm p}$ of the initial
N--body model is somewhat too large.

\subsection{Best model}

\noindent
As we have argued, the best values for the free parameters
are given by the 44\degr\ solution of {\bf bd} : {\bf bd1}.
From the free parameters we can derive some more interesting
properties of the best--fitting model. Corotation radius is
at 4.5 kpc and the semi--major axis of the Bar is 2.5 kpc 
(cf. Section 3.1). The pattern speed is 46 \kmsr\ for
the rescaled local circular velocity of 207 \kms . Finally, the mass of
the Bar is 1.7$\times10^{10}$ \msun .
The plausibility of {\bf bd1} is 15$\pm$7\% ,
but when determining $\langle P\rangle$ 
with only the double--peaked stars ({\bf bdd}) we
get 68$\pm$8\% .  The single--peaked stars in {\bf bd},
that possibly do not (all) belong to the OH/IR (asymptotic--giant
branch) star population, apparently decrease the goodness of the
fit.

\section{Discussion}

\beginfigure7
\fignam\PROFS
{\psfig{figure=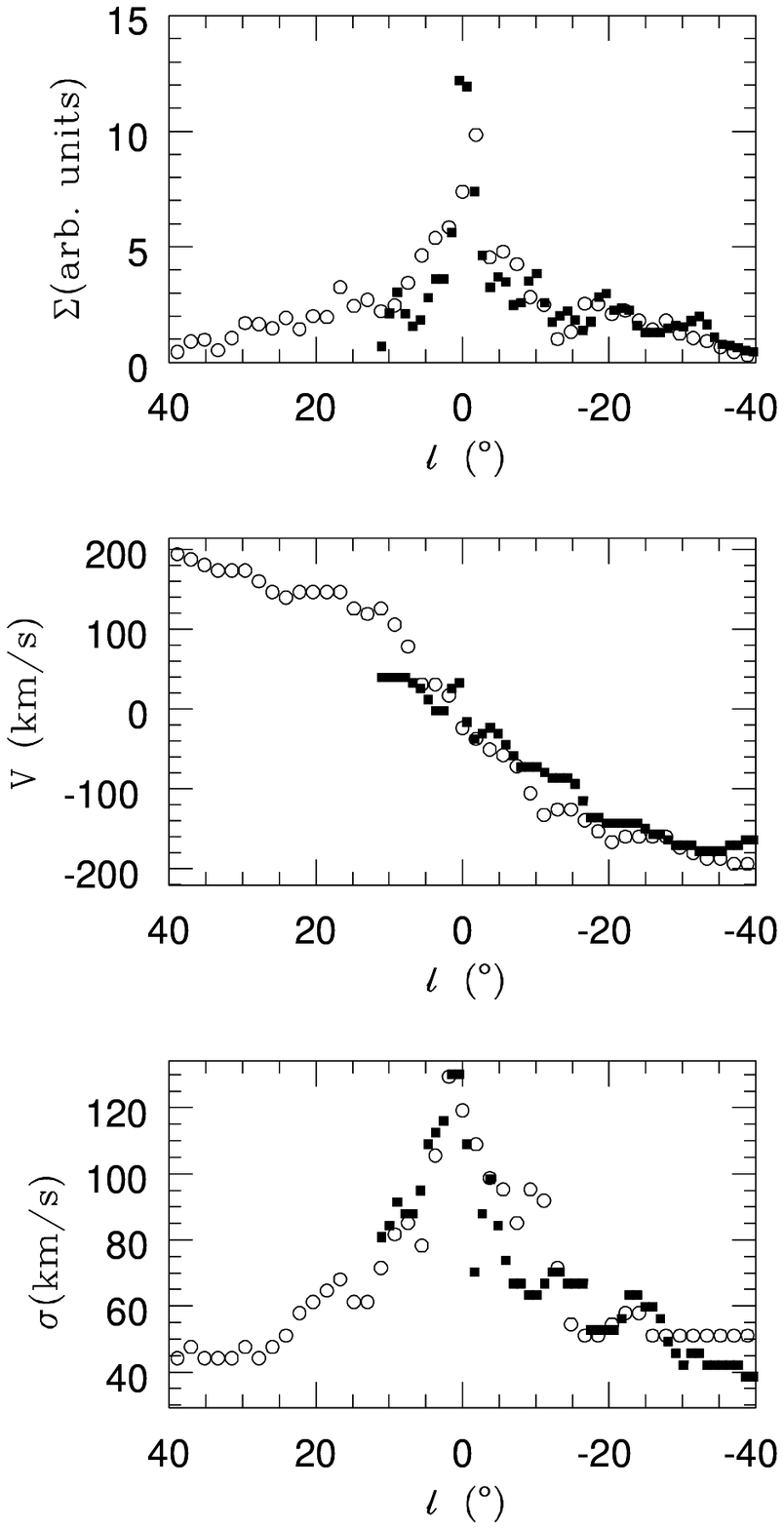,width=8cm}}
\caption{{\bf Figure \nfig}
Surface density, 
mean galactocentric \losa\ velocity and dispersion 
(all at $b=0^{\circ}$) for
model {\bf bd1} in open circles and for data (total
OH/IR sample) in filled squares.
The points shown are obtained via 3--D--adaptive--kernel smoothing.
Note that hence the distributions
are not reliable at the borders of the observed region ($\ell$\gsim 7\degr;
$\ell$\lsim $-$42\degr).
}
\endfigure

\noindent
The value of 44\degr\ we obtain for $\phi$ 
is large compared to some other estimates 
(16\degr, Binney \etal1991 (gas dynamics); 20\degr, 
Binney, Gerhard \& Spergel 1997 (integrated light); 24\degr,
Nikolaev \& Weinberg 1997 (star counts); 20\degr --30\degr, 
Stanek \etal1997 (flux differences between positive and negative
longitudes)).
It should be noted that
the lowest value for the viewing angle, 16\degr\ (Binney \etal1991),
is based on the possibly wrong assumption that
the CO ``parallelogram'' (Bally \etal1988) is formed by gas on the
inner cusped orbit (see \CHSIX). The parallelogram
may well be the result of a tilt in the inner gas 
disk (Liszt \& Burton 1978).
Our high $\phi$ is compatible with the COBE E2\&E3 models ($\phi\sim$40\degr) 
by Dwek \etal(1995) and with the value of 35\degr\ 
determined by Weiner \& Sellwood (1996; from the gas dynamics).
Also, Unavane \& Gilmore (1998) find from near--infrared
starcounts that models with viewing angles between 20\degr --45\degr\ are
acceptable. 
A much older determination of $\phi$ from the gas dynamics even
yields 45\degr\ (Peters 1975).
To explain the two local star streams, Kalnajs (1997)
argues that the viewing angle should be 45\degr\ as well.

Not only the high value for viewing angle, but also the
value for the corotation radius, or pattern speed, can be 
reconciled with the observed gas kinematics. Various
attempts to model HI and/or CO kinematics in the inner 4 kpc 
have given values for the pattern speed ranging from 
\pspeed$\sim$19\kmsr\ (Wada \etal1996) to
63\kmsr\ (Binney \etal1991) or even 118\kmsr\ (Yuan 1984).
Hydrodynamic models for the HI \lvd\ of the whole Galaxy
are illustrated in Mulder \& Liem (1986),
who themselves give as best model parameter $\phi$=20\degr\ and
$R_{\rm CR} \sim $\rsun\ (their Fig.$\,$5), but 
their model with $\phi$=40\degr\ and
$R_{\rm CR} \sim $0.5\rsun\ (their Fig.$\,$9) gives a 
similarly good and arguably better
fit to the 3--kpc arm and the central CO kinematics (Bally \etal1988).

In Fig.$\,$\PROFS\  we illustrate how the data compare
to the {\bf bd1} model. The global agreement
is good and also the region around $\ell=-$20\degr\ which 
may be in the corotation region of the Galaxy 
(Sevenster 1997) shows very similar
features in all three quantities in model and data.

Fux himself (1997) derives a
best--fit value for $\phi$ of 25\degr\ 
with the same N--body model (his values for
the other parameters are as in {\bf fixf}, Table \BESTFF).
We find that the values 
for $\phi$ found in run {\bf fixf} are no different from 
the other runs.
To mimic the model optimization from only the COBE K--band 
surface density (Fux 1997; his value \fv\ comes from
scaling the velocities to fit the \losa\ dispersion 
towards Baade's window) we introduced run {\bf morx} where
the kinematics are `switched off' completely by setting
$N_{\rm V}$=1. Indeed, one of the solutions gives
$\phi$=29\degr\ (and \rsun =9.0 kpc), but it carries zero plausibility.
As argued before, the global stellar kinematics
seem essential in determining the viewing angle as they 
provide the necessary constraint to suppress the degenerating
influence of limited windows and distance coverage.
In Fux's (1997) case the longitude window ($+$30\degr, $-$30\degr)
is large enough to prevent the problems we discussed in Section 4.1.
However, latitudes $|b|<$3\degr\ are excluded from the optimization
because the COBE data cannot be corrected reliably for
extinction in the plane. This reminds us of run {\bf ptl} 
with its incompleteness in the plane and subsequent low viewing angles and 
incapability of constraining the model ($\langle P\rangle$=0\%).
We would argue that many of the $\phi\sim$25\degr --30\degr\ results
found in the literature suffer from similar problems.

The measure for the relative residual defined by Fux, $R^2_{N_{\rm pix}=300}$,
is 1.5\% for {\bf bd1} (cf. 0.47\% for his best fit for m08t3200).

It has proven virtually impossible to distinguish between 
triaxial and axisymmetric distributions studying the 
distribution of the 
\losa\ velocities only (eg. Ibata \& Gilmore 1995; Dejonghe \etal1997).
We applied the Kolmogorov--Smirnov test as used by 
Ibata \& Gilmore (1995), as well as the intrinsically 
more powerful distance--velocity
statistic described by Dejonghe \etal(1997), 
to the N--body model. 
Both only give significant results
for very low latitudes ($ |b| < 2^{\circ}$) and even then only
with samples of at least 1000 stars. 
The results are very dependent on the viewing angle.
It is therefore
no surprise that Ibata \& Gilmore (1995) found no significant
evidence for triaxiality from their intermediate--latitude
velocity profiles and that neither statistic gives a signal
when applied to the AOSP sample.
However, even though the relatively low surface density of the AOSP sample 
inhibits the construction of velocity
profiles, it contains sufficient kinematic
evidence of triaxiality, as indicated by the plausibility of $>$50\% . 
The low latitudes, very homogeneous sampling and simultaneous
fitting of the spatial and the kinematical distribution 
are essential.

In an earlier stage of this project, we applied the
same procedure to the Schwarzschild--type
N--body model by Zhao (1996) in its 
initial state before evolution. No significant fit
was obtained for this unmixed model.
Fux's N--body model has proven to have a solid
physical basis and shows many observed features in a variety
of data (Fux 1997; Fux \& Friedli 1996). 
Also the N--body bar's formation, spontaneously via 
instability of the underlying disk,
is one of the probable ways to form bars (eg. Sellwood \& Wilkinson 1993).
The fact that for the {\bf ptl} sample, selected especially for
its incompleteness (see Section 3.2), no significant
fits can be obtained, gives extra credibility to the model (as 
well as the method). This all provides
proof that m08t3200, Fux's (1997) best model, gives
a very good representation of the six--dimensional Galaxy.

Clearly, there is room for improvement, especially to lift
the $\phi$--degeneracy once and for all. The most important features
of a stellar data set used to achieve this are a large 
longitude range and homogeneous sampling of low latitudes.
More dimensions to limit the degrees of freedom
are preferred over more objects.

\section{Micro--lensing optical depth in the line of sight to the Bar}

\beginfigure8
\fignam\TAU
{\psfig{figure=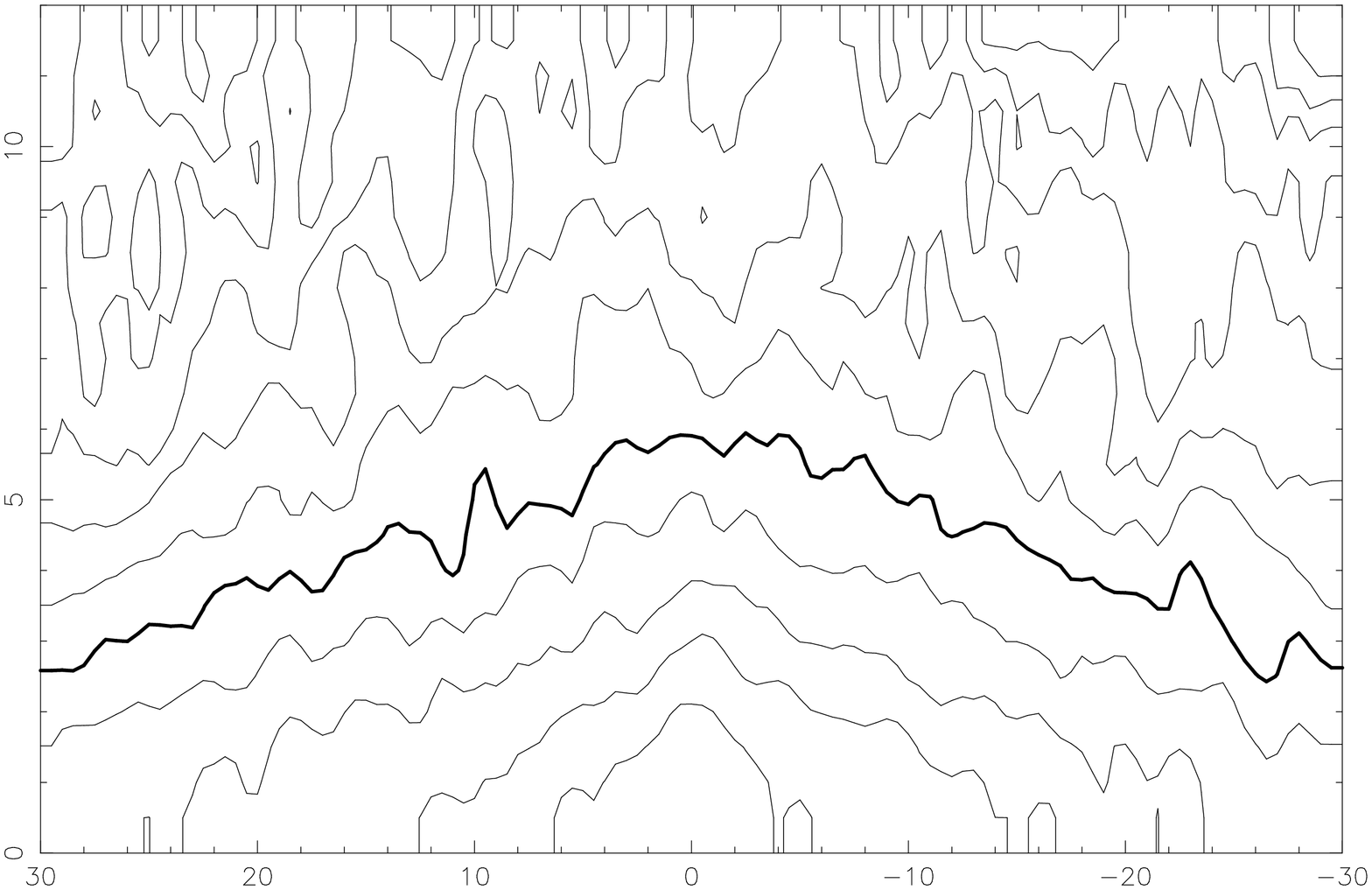,angle=270,width=8.5truecm}}
\caption{{\bf Figure \nfig }
Micro--lensing--optical--depth ($\tau_0$) map, symmetrized
in latitude, for {\bf bd1}, without the contribution of the
dark particles in the simulation (see Fux 1997).
The thick contour is for $\tau_0=1.0\times 10^{-6}$; the other
contours are spaced by a factor of 1.5, decreasing with increasing
latitude.
}
\endfigure

\noindent
In this section, we want to discuss briefly the micro--lensing properties
of the scaled model. The observed micro--lensing optical depth
is as yet unaccounted for by any density model for the central Galaxy
so it is important to assess this optical depth for our best model.
In a forthcoming paper, we will do this in more detail and 
also calculate the event--duration distribution, that may give
insight in the nature of the missing optical depth, even though
the stellar mass function is not known in the model.

The micro--lensing optical depth $\tau$ is the probability to
detect a micro--lensing event at a given instant. From a
density model, one can calculate the distribution of $\tau$
on the sky; the micro--lensing--optical--depth map.
Comparison with the measured values 
($\tau=3.9^{+1.8}_{-1.2}\times 10^{-6}$ for red clump giants
toward ($\ell$ = \decdeg 2.55, $ b=-$\decdeg3.64);
$\tau=2.1^{+0.5}_{-0.4}\times 10^{-6}$ for main--sequence stars,
corrected for blending,
toward ($\ell$ = \decdeg 2.7, $ b=-$\decdeg4.1),
Alcock \etal1997; $\tau=3.3^{+1.2}_{-1.2}\times 10^{-6}$ toward Baade's window
($\ell \sim $1\degr, $ b\sim$ -4\degr), Udalski \etal1994)
gives important information about the model.
For a wide range of models that derive from the COBE
maps (Dwek \etal1995),
one finds that $\tau_{\rm mod}$ to be 2$\sigma$ lower than $\tau_{\rm obs}$ 
(Zhao \& Mao 1996).
The missing optical depth thus has to be accounted for,
within the limits put by other observations, by
a component not present in those models; either
dark or sub--stellar particles or an extra
density component (eg. a thick disk).

For the calculation of $\tau$ one needs to take the brightness
of the lensed sources into account. This is some function of 
their distance $D_s$, so that 
the optical depth in a certain direction also depends on $D_s$.
Kiraga \& Paczynski (1994) hence defined 
$\tau_{\beta} \propto D_{s}^{2+2\beta}$, where $\beta$ defines
the exact dependence on $D_s$ and from the subscript of $\tau$
one can immediately see which dependence was used in the 
theoretical calculation. For $\beta=0$, the sources are visible
out to infinity; we get $\tau_{0} \propto D_{s}^2$ and the proportionality
is a straightforward volume correction.

For {\bf bd1}, we calculated the $\tau_0$ map (Fig.$\,$\TAU), 
as described in Fux (1997). 
It is well known that only a 
strong and massive bar, with a viewing angle smaller
than 20\degr\ (Zhao \& Mao 1996; Kiraga \& Paczynski 1994;
Zhao 1997a; Fux 1997 his Fig.$\,$13),
can fully account for the measured values.
Models derived from observed surface--density measurements
have bars too weak, and mostly at too high viewing angle, 
to be able to account for $\tau$ (eg. Nikolaev \& Weinberg 1997).
The values we obtain in the direction of the measurements
are $\tau_0 \sim 1.5\times 10^{-6}$, best compared to the
$3.9\times 10^{-6}$ value for the clump giants, and
$\tau_{-1} \sim 0.9\times 10^{-6}$, best compared to
$2.1\times 10^{-6}$.
Indeed both $\tau_0$ and $\tau_{-1}$ (both without the contribution 
of the dark component) are too low to explain observations
by $>2\sigma$, thus confirming the preliminary results presented by
Valls--Gabaud \etal(1997).

The distribution of the micro--lensing optical depth is not very
bulge--like; it is dominated by lenses in the disk. There is
no significant asymmetry between positive and negative longitudes,
as would be expected when either lenses or sources (or both) have
a barred distribution (see Evans 1994). 
However, the asymmetries in the optical--depth distribution
from disk lenses and bar lenses, respectively, have opposite
signs (Evans 1994). Simple tests, with $\beta=0, -1$,
show that disk lenses skew the distribution 
toward positive longitudes; bar lenses toward negative longitudes.
The net effect clearly depends on $\beta$, $\phi$ and density parameters.

In fact, the asymmetric signal is expected to be largest
for a viewing angle 45\degr\ (eg. Evans 1994). 
In principle, since in Fig.$\,$\TAU\ we show the $\tau_0$ map ($\beta=0$),
we expect the distribution to be skewed toward negative longitudes.
Additional spiral arms, protruding from the ends
of the Bar (see Fux 1997, m08t3200 model), could
counteract this asymmetry. 
All in all, the symmetric appearance of Fig.$\,$\TAU\ is understandable.

\section{Conclusions}

We presented values for the free parameters of a self--consistent
model of the Galaxy, optimized to fit positions and velocities
of various sets of evolved stars. These stars 
are representative for the global galactic distribution.
The method is found to be sensitive to 
incompletenesses and to large--scale kinematics.

The over--all distribution of the stars 
is fitted well by a bar of the global form of that of the N--body
model (Fux 1997) with a semi--major axis of 2.5 kpc,
corotation radius of 4.5 kpc,
an axis ratio of 0.5 and a viewing angle of 44\degr.
The value for the viewing angle is high but not incompatible
with previous determinations from stellar data
as well as gas kinematics. The mass contained within
this Bar is $\sim 1.7\times 10^{10}$\msun, marginally lower than
various other derivations (Zhao \etal1996; 
Blum 1995; Kent 1992). The derived pattern speed for the Bar, 
$\Omega_p = V_c(R_{\rm CR})/R_{\rm CR}$ is 
46 \kmsr.
For the solar azimuthal velocity a low value of 171 \kms\ 
is found. This is much lower
than the local circular velocity (207 \kms) in the model potential, implying
that the Sun (more precise, the local standard of rest) 
would not be on a circular orbit in this particular model.
This is in agreement with the findings of Kuijken \& Tremaine (1994)
who find a local circular velocity of 200 \kms\ and \vsun = 180 \kms.

We argue that using low--latitude, unbiased, global stellar
kinematics is crucial to determine the viewing angle $\phi$.
The commonly found and accepted low values 
of $\phi\sim$ 25\degr\ (see Section 5) should be viewed with caution.
Our method, applied to a variety of data sets with `known 
flaws' or to the stellar positions only, shows that 
indeed these favour viewing angles around the lower value.
The resulting fits do not have high plausibility.

One of the reasons is that
the signal of the Bar diminishes quickly
with increasing latitude and thus $\phi$ becomes
ill constrained. The sample of {\bf ptl},
that is underrepresenting the plane below $\sim$3\degr\ -- approximately
one scaleheight -- gives a null--result. 
This means that either there has not been significant
bar--induced thickening in the inner Galaxy, or the thickening
conspires with the distribution becoming rounder.
We believe that parameters of the \gba\ can not be reliably constrained 
without data that trace its inner one scaleheight.

This also means that Baade's window may not be an appropriate
region to sample the Bar's properties (see also \CHSIX).
Regardless of this consideration,
with the best--model viewing angle of 45\degr\
the Bar does not give significantly higher values
for $\tau$ than do a variety of axisymmetric distributions
(Zhao 1997a; Kuijken 1997) and $\tau_{\rm mod}\sim0.5 \tau_{\rm obs}$
for our best model.
We conclude that (provided that the value of $\tau_{\rm obs}$ is beyond
suspicion) the origin of the discrepancy between current bar models
and the observed micro--lensing optical depth should be
sought in a foreground component - eq.~a spiral arm or a 
thick disk - with larger apparent scaleheight than the Bar.
The most convincing argument in favour of a small
viewing angle for the Bar (Zhao 1997a) is thus taken away.

\section*{Acknowledgments} 
MNS thanks Mt.~Stromlo and Siding Spring Observatories for
hospitality during the start of this project.
The visit was financed from an Amelia Earhart
Fellowship granted by Zonta International Foundation.
DVG also thanks the warm hospitality of MSSSO,
and the French--Australian
Committee for Astronomy for travel support.
Useful suggestions and hints by
Tim de Zeeuw, Adriaan Blaauw and the referee David Spergel
are acknowledged.



\section*{References}
\beginrefs

\bibitem Alcock C. \etal1997\apj 479 119
\bibitem Bally J., Stark A., Wilson R., Henkel C. 1988\apj 324 223
\bibitem Baud B., Habing H., Matthews H., O'Sullivan J., Winnberg A.,
      1975\nature 258 406

\bibitem Binney, J.J., Gerhard, O.E., Stark, A.A., Bally, J., Uchida, K.I.,
 1991\mnras 252 210
\bibitem Binney J.J., Gerhard O.E., Spergel D.N., 1997 \mnras 288 365

\bibitem Blum R., 1995\apjl 444 89
\bibitem Charbonneau P., 1995\apjs 101 309

\bibitem Dehnen W., Binney J., 1998\mnras 294 429
\bibitem Dejonghe H., Vauterin P., Caelenberg van K., Durand S., Mathieu A.,
    1997, IAU Symp.180, Planetary Nebulae, p.428
\bibitem Dwek E. \etal1995\apj 445 716

\bibitem Elmegreen B.\bargal 197
\bibitem Evans N.W., 1994\apjl 437 31
\bibitem Feast M,. Whitelock P., 1997\mnras 291 683
\bibitem Frogel J., 1988\araa 26 51

\bibitem Fux R., 1997\aa 327 983
\bibitem Fux R., Friedli D. \bargal 529
\bibitem Gerhard O.\solve 79
\bibitem Habing H.J.\gents 57
\bibitem Ibata R., Gilmore G., 1995\mnras 275 605
\bibitem Kalnajs A.J., 1997, In: Sandqvist, Lindblad (eds),
Barred Galaxies and Circumstellar Activity, Lect.n. in Phys. 474, p.165

\bibitem Kent S., 1992\apj 387 181
\bibitem Kiraga M., Paczynski B., 1994\apjl 430 101
\bibitem Kuijken K., Tremaine S., 1994\apj 421 178
\bibitem Kuijken K., 1997\apjl 486 19
\bibitem Liszt H., Burton W., 1978\apj 226 790 

\bibitem Mulder W., Liem B., 1986\aa 157 148

\bibitem Nikolaev S., Weinberg M., 1997\apj 487 885
\bibitem Noguchi M.\bargal 339
\bibitem Peters W., 1975\apj 195 617


\bibitem Press W., Teukolsky S., Vetterling W., Flannery B., 1992, 
  ``Numerical Recipes'' , Cambridge University Press

\bibitem Rohlfs K., Boehme R., Chini R., Wink J., 1986\aa 158 181

\bibitem Saha P., 1998\aj 115 1206

\bibitem Schwarzschild M., 1979\apj 232 236
\bibitem Sellwood J., Wilkinson A., 1993, Rep. Prog. Phys., 56, 173

\bibitem Sevenster M.N., Chapman J.M., Habing H.J., Killeen N.E.B., Lindqvist M., 1997a\aas 122 79 (S97A)

\bibitem Sevenster M.N., Chapman J.M., Habing H.J., Killeen N.E.B., Lindqvist M., 1997b\aas 124 509 (S97B)

\bibitem Sevenster M.N., 1997, dissertation Leiden University 
\bibitem Stanek K. \etal1997\apj 477 163
\bibitem te Lintel Hekkert P., Caswell J., Habing H.J., Haynes R., Wiertz W., 
 1991\aas 90 327

\bibitem Udalski A. \etal1994, Acta Astron. 44 165
\bibitem Unavane M., Gilmore G., 1998\mnras 295 145
\bibitem Valls--Gabaud D., Saha P., Sevenster M., Fux R., 1997\varmic 119
\bibitem van Langevelde H., Brown A., Lindvist M., Habing H., de Zeeuw P.,
  1992\aal 261 17

\bibitem Veen, W. van der, 1989\aa 210 127
\bibitem Wada K., Taniguchi Y., Habe A., Hasegawa T.\bargal 554
\bibitem Weiner B., Sellwood J.\solve 145
\bibitem Yuan C., 1984\apj 281 600
\bibitem Zhao H.S.,  1996\mnras 283 149
\bibitem Zhao H.S., Mao S., 1996\mnras 283 1197
\bibitem Zhao H.S., Rich R.M., Spergel D., 1996\mnras 282 175
\bibitem Zhao H.S.,  1997a\varmic 109
\bibitem Zhao H.S.,  1997b, astro--ph 9705046
\bye